\documentclass[%
preprint,
 amsmath,amssymb,
 aps,
]{revtex4-2}

\usepackage{graphicx}
\usepackage{dcolumn}
\usepackage{bm}
\usepackage{hyperref}
\hypersetup{
    colorlinks=true,
    linkcolor=blue,
    }

\usepackage{gensymb}
\usepackage{todonotes}

\begin{document}

\preprint{APS/123-QED}

\title{Femtosecond-Terawatt Hard X Ray Pulse Generation with Chirped Pulse Amplification on a Free Electron Laser}

\author{Haoyuan Li}
\altaffiliation{
Also at Physics Department, Stanford University, Stanford, California, 94305, U.S.A.
}
\author{James MacArthur}
\author{Sean Littleton}
\altaffiliation{
Also at Applied Physics Department, Stanford University, Stanford, California, 94305, U.S.A.}
\author{Mike Dunne}
\author{Zhirong Huang}
\altaffiliation{
Also at Applied Physics Department, Stanford University, Stanford, California, 94305, U.S.A.}
\author{Diling Zhu}
\email{dlzhu@slac.stanford.edu}
\affiliation{
Linac Coherent Light Source, SLAC National Accelerator Laboratory, Menlo Park, California, 94025, U.S.A.
}

\date{\today}

\begin{abstract}

Advances of high intensity lasers have opened up the field of strong field physics and led to a broad range of technological applications. Recent x ray laser sources and optics development makes it possible to obtain extremely high intensity and brightness at x ray wavelengths. In this paper, we present a system design that implements chirped pulse amplification for hard x ray free electron lasers. Numerical modeling with realistic experimental parameters show that near-transform-limit single-femtosecond hard x ray laser pulses with peak power exceeding 1 TW and brightness exceeding $4\times10^{35}~$s$^{-1}$mm$^{-2}$mrad$^{-2}$0.1\%bandwdith$^{-1}$ can be consistently generated. Realization of such beam qualities is essential for establishing systematic and quantitative understanding of strong field x-ray physics and nonlinear x ray optics phenomena.
\end{abstract}

\keywords{chirped pulse amplification, free electron laser, hard x ray, terawatt}
\maketitle

High-intensity high-brightness x-ray pulses from free electron laser (FEL) have opened up many new routes of research for strong field physics \cite{fuchs2015anomalous, krebs2019time}, non-linear x-ray optics, and many potential applications \cite{PhysRevLett.112.163901, PhysRevLett.120.223902, PhysRevLett.120.023901, schori2017parametric, sofer2019observation, krebs2021theory, kroll2018stimulated, glover2012x, marcus2014free, PhysRevB.72.235110}.
While the past decade have witnessed many \emph{first demonstrations}, one major obstacle towards a quantitative, systematic and application-oriented understanding of the subject came from the stochastic temporal and spectral structure of the pulses \cite{saldin1998statistical}.
Production of Terawatt-femtosecond (TW-fs) hard x ray pulses with full spatial and temporal coherence is therefore highly desired.
This calls for allround-improvement of the x ray beam quality, including peak intensity, peak brightness, and a well defined spatial temporal profile, which are mandatory for quantitative analysis and prediction of the nonlinear observables.

Currently, the peak power of a state-of-art free electron laser (FEL) can reach 100~GW scale \cite{PhysRevLett.120.014801} through self-amplified spontaneous emission (SASE).
To push the peak power into TW scale, several enhanced-SASE schemes \cite{PhysRevSTAB.8.040701, kumar2016temporally, shim2018isolated} have projected terawatt-attosecond (TW-as) output by using extremely high peak current.
A super-radiance based approach \cite{prat2015simple} was shown to be capable of delivering TW hard x-ray pulses with a sequence of short electron bunches and electron and optical delay devices.
These approaches still inherited the rugged temporal characteristics of SASE, while also requiring electron beam parameters beyond the current state-of-art.
Self-seeding techniques were adopted to improve the temporal coherence up on SASE \cite{amann2012demonstration,inoue2019generation,min2019hard, osti_1560969} with impressive peak brightness improvement.
However, the narrow bandwidth puts a limit on attainable pulse duration and peak power.
In this letter, we present the design and performance evaluation of a hard x-ray chirped pulse amplification (CPA) setup, that can potentially deliver single-femtosecond x-ray pulses with a clean temporal profile and provide high peak power ($>1$ TW) and high peak spectral brightness at the same time.

\begin{figure*}[bht!]
    \centering
    \includegraphics[width=\textwidth]{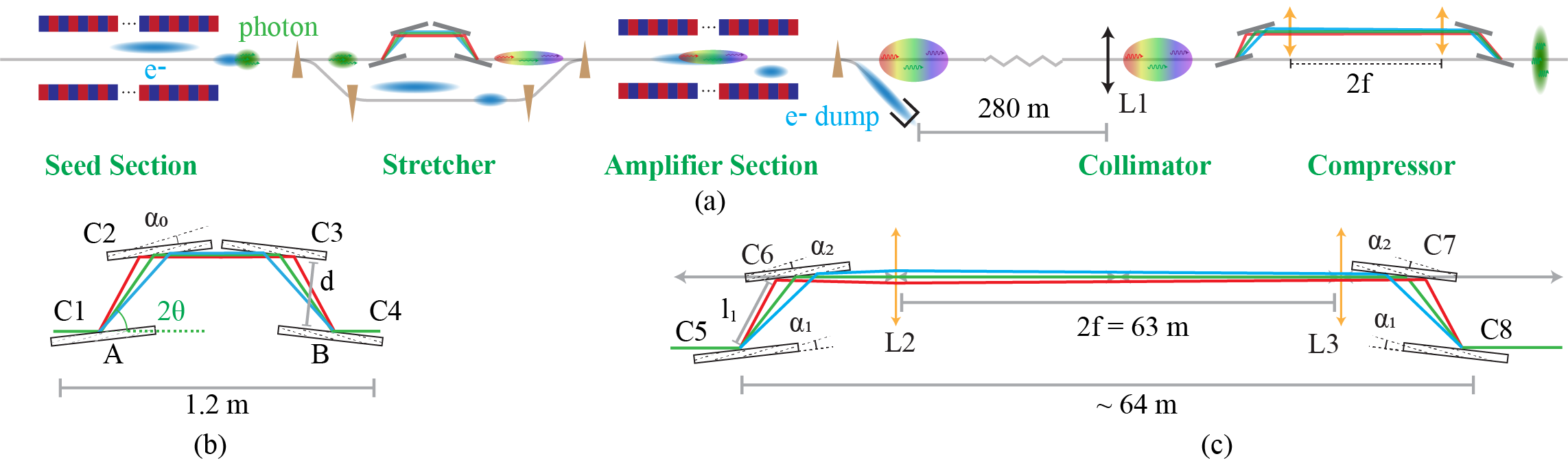}
    \caption{
    (a) Layout of the setup. 
    The symbols $C1$ to $C8$ represent Bragg crystals to guide the trajectory of the x ray pulse.
    (b) Schematics of the stretcher.
    (c) Schematics of the compressor.
    }
    \label{fig:layout}
\end{figure*}

At optical wavelength, CPA has revolutionized high intensity pulse generation and its applications \cite{strickland1985compression, papadopoulos2016apollon}.
The concept has been extended to the extreme-ultraviolet wavelengths\cite{gauthier2016chirped}. 
At \AA ~wavelength, an earlier proposal \cite{pellegrini2000high} used chirped electron bunches to generate chirped x ray pulses before the compression by an optical compressor.
Detailed analysis \cite{krinsky2003frequency} showed that, however, electron bunch manipulation alone cannot produce temporal chirp of sufficient magnitude and quality, leading to limited compression ratio and low photon throughput at the compressor.
We overcome this limitation by a complete CPA scheme consists of both optical stretcher and compressor.
This gave us precise control of the temporal chirp, realizing two seemingly competing quantities in an FEL: a relatively wide bandwidth required by the short final pulse duration and the long and uniform pulse temporal profile during amplification to extract as much energy from the electron bunch as possible, simultaneously. The system design is illustrated schematically in Figure~\ref{fig:layout} (a), where bold green labels are used to indicate the name of different parts.

In this hypothetical setup, the undulator segments are divided into two groups: the seed section and the amplifier section.
We use two electron bunches with nanosecond time separation \cite{decker2022tunable} to serve as the seed and amplification gain media respectively.
The seed electron bunch is short with high peak current.
The amplifier electron bunch is longer and contains higher total charge.
In the seed section, the first electron bunch produces a short and intense hard x ray pulse while the trailing amplifier electron bunch travels along a lasing-suppressed orbit.
Exiting the seed section, the seed pulse propagates through a crystal-optic stretcher, and obtains a strong temporal chirp while expanding significantly in pulse duration, matching that of the amplifier electron bunch.
The electron bunches are diverted away from the stretcher optics after the seed section with a magnetic chicane.
The crystal stretcher introduces a time delay to the seed x ray pulse by the exact amount as the time spacing between the seed and amplifier electron bunches.
This enables interaction between the stretched seed x ray pulse and the second electron bunch to achieve amplification in the amplifier section of the undulator, where the seed electron bunch is directed to a lasing-suppressed orbit.
Exiting the amplifier section, both electron bunches are diverted to the electron beam dump.
The amplified, saturated, and chirped hard x ray pulse propagates downstream to a crystal optic compressor, where we remove the temporal chirp to reach a much shorter pulse duration and higher peak power.

To understand the potential performance envelop of this CPA scheme, detailed numerical modeling is required.
The chosen electron beam parameters for the seed and amplifier bunches are summarized in Table~\ref{table:parameters}, all of which have been demonstrated at many current FEL facilities, such as LCLS.
A total of 32 undulator segments as currently installed at the LCLS-II hard x-ray beamline were assumed.
We used \emph{Genesis} \cite{reiche1999genesis} to predict the electron lasing dynamics in both undulator sections.
The crystal optics is simulated with a homebuild beam propagation program, the source code of which can be found in Ref \cite{li_2022}.
The 2-beam dynamical diffraction theory \cite{batterman1964dynamical} is used to describe the reflection, angular dispersion and absorption of crystal optics and transmission optics.
The operation photon energy is chosen to be 9.5~keV.

\begin{table}[hbt!]
    \begin{tabular}{l|l|l}
    \hline
                    & $1^{st}$ electron bunch       & $2^{nd}$ electron bunch         \\ \hline \hline
    Bunch Charge    & 7.5 pC   & 200 pC     \\ \hline
    Bunch Length    & 0.375 um & 20 um      \\ \hline
    Emittance       & 0.4~um   & 0.4~um   \\ \hline
     Energy Spread  & 3.49~GeV & 2.06~MeV   \\ \hline
    Current Profile & Flat-top & Flat-top         \\ \hline
    Peak Current    & 6 kA     & 3 kA             \\ \hline
    Taper           & Flat     & Genesis-informed \\ \hline
    K               & 2.4      & 2.4              \\ \hline
    Electron Energy & 10 GeV   & 10 GeV           \\ \hline
    \end{tabular}
    \caption{Beam parameters used in \emph{Genesis} simulations.}
    \label{table:parameters}
\end{table}

The schematics of the stretcher is shown in Figure~\ref{fig:layout} (b).
This is a direct analog of the corresponding device at optical wavelengths \cite{martinez1984negative}.
The stretcher consists of 4 crystals, $C1$ to $C4$, arranged in a mirror symmetric layout, with asymmetric Bragg reflections.
Due to the angular dispersion of asymmetric Bragg reflections, x ray photons with different energies follow different trajectories inside the stretcher, indicated by different colors in Figure~\ref{fig:layout} (b), which leads to a light path length difference.
Define $\theta$ to be the Bragg angle, $\alpha_0$ the asymmetry angle of the Bragg reflection, $d$ the gap size of the crystal pairs $C1$/$C2$ and $C3$/$C4$, and $E_0$ the reference photon energy.
With ray-tracing analysis, the energy dependence of the path length inside the stretcher can be shown to be
\begin{equation}
    \delta (L)\approx-8d\frac{\sin^2{\alpha_0}\sin^2{\theta}}{\sin^3{\left(\alpha_0+\theta\right)}}\times\frac{\delta E}{E_0},
    \label{eq:stretcherPL}
\end{equation}
where $\delta E$ is the energy difference with respect to $E_0$.
Note that higher energy x rays always have a shorter path length.
This thus always creates a negative temporal chirp regardless of the sign of $\alpha_0$.

A large bandwidth and high reflectivity are desired, in order to obtain higher seed pulse energy, and reduce potential compressed pulse duration.
We choose Si (111) reflections for $C1$ to $C4$, with an asymmetry angle of $10\degree$ to produce the required angular dispersion and increase the bandwidth.
At $E_0=9.5$~keV, the Bragg angle is $\theta = 12.02\degree$.
As shown in Figure~\ref{fig:layout} (b), the beam adopts grazing incidence geometry on $C_1$/$C3$, and grazing exit geometry on $C2$/$C4$.
The gap size $d$ must be chosen to match the stretcher delay to the time separation between the two electron bunches, which can only be multiples of 0.35~ns, determined by of the radio-frequency (RF) of the accelerator \cite{decker2015two}.
Following geometric optics, the relation between stretcher delay $T_{\text{delay}}$ and gap size $d$ can be shown to be
\begin{equation}
    T_{\text{delay}} = \frac{2d}{c}\frac{\left(1- \cos{2\theta }\right)}{\sin{(\theta-\alpha_0)}}  .
\end{equation}
where $c$ is the speed of light in vacuum.
Therefore, the available gap sizes are $d\in\{2.1~\text{mm}, 4.3~\text{mm}, \cdots\}$.
The $d=2.1~\text{mm}$ is chosen in order to match the stretched seed pulse duration to the amplifier electron bunch length, which is limited to a maximum of $\sim$60~fs if we want to maintain the optimal peak current of 3~kA.
At $d=2.1~\text{mm}$, simulation of 60 seed SASE pulses with a FWHM pulse duration of $0.6\pm 0.6~\text{fs}$ at $9.5~\text{keV}$ yielded FWHM pulse duration after the stretcher at $43.8\pm 11.3~\text{fs}$.

\begin{figure}[bht!]
    \centering
    \includegraphics[width=0.47\textwidth]{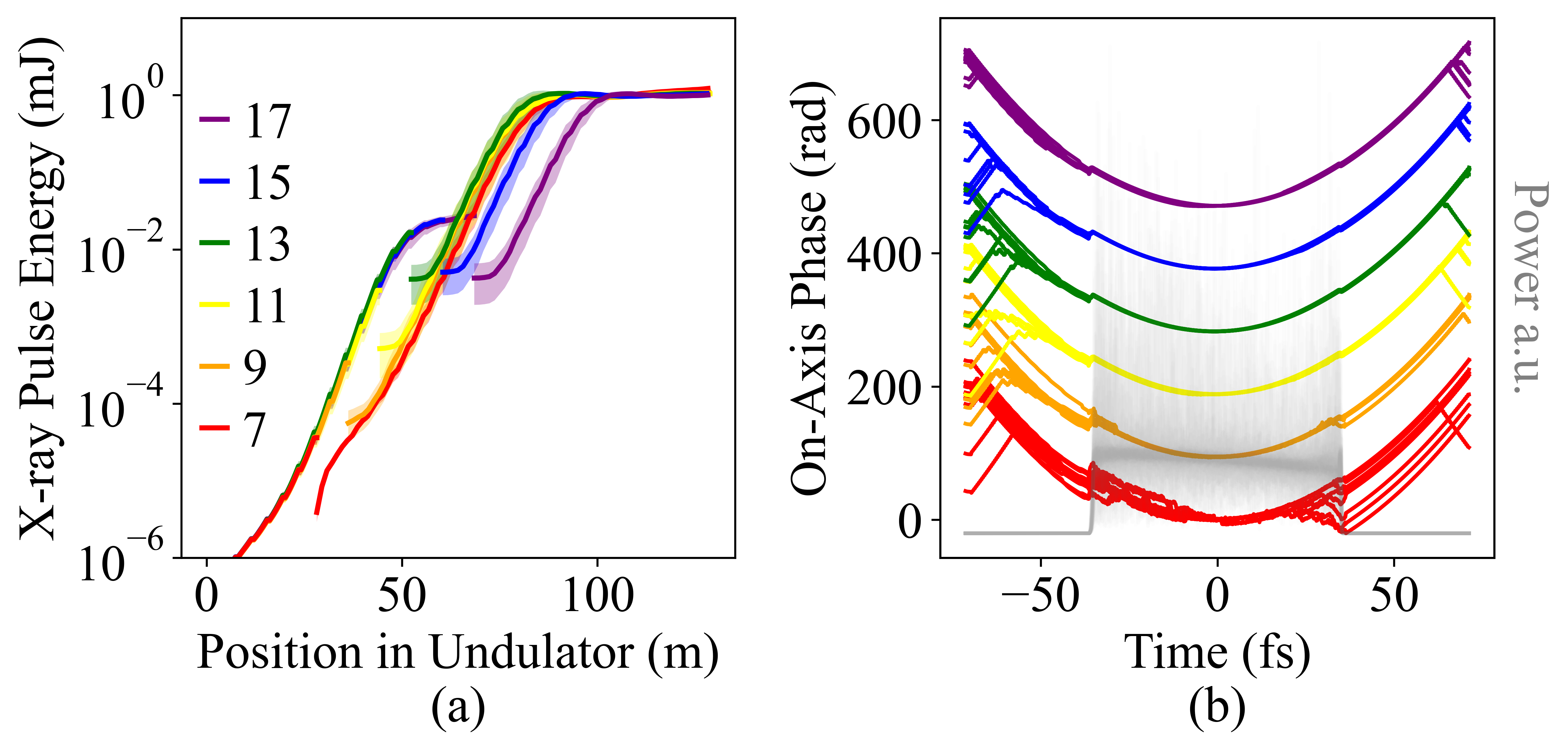}
    \caption{(a) The evolution of the x ray pulse energy along the undulator sections for different configurations. 
    The number in the legend is the number of undulator segments assigned to the seed section.
    The solid lines represent the averaged pulse energy over 10 simulated pulses in each configuration.
    The shaded area indicated the standard deviation of the pulse energy of the 10 pulses.
    (b) The on-axis phase of x ray pulses at the exit of the amplifier section.
    Each solid line represents the phase of a single simulated pulse.
    The color scheme and its meaning is the same as that of subplot (a).
    The grey shadow indicate the power profile of all 70 pulses. 
    }
    \label{fig:undulator}
\end{figure}

Using the stretcher parameters specified above, we next optimize the distribution of undulator segments between the seed section and amplifier section.
In total, 6 configurations are compared, from 7 to 17 undulator segments assigned to the seed section with increments of 2.
The rest undulator segments are assigned to the amplifier section.
For each configuration, 10 simulations of the FEL pulse energy growth trajectory are summarized in Figure~\ref{fig:undulator} (a), with different colors representing different undulator allocation schemes.
One can see that when the number of seed undulators reaches 13, the pulse energy of the seed pulse after the stretcher saturates.
Allocating additional seed segments would not increase the seed power but rather would impose higher thermal load to the first crystal of the stretcher.
The on-axis phase of each pulse as shown in Figure~\ref{fig:undulator} (b), one sees the prominent parabolic phase curves across all settings, indicating that the electron dynamics in the amplifier section preserves the temporal chirp very well except for the 7-seed-segment case, where the seed power is too low.
The compressor has a similar bandwidth as the the stretcher.
Therefore the weak parasitic SASE radiation will be filtered out by the compressor.
Reconciling the desire for a strong seed pulse, low crystal thermal load, and the preservation for temporal chirp, we choose the configuration with 13 seed undulators.
We then apply a \emph{Genesis}-informed taper optimization for this configuration.
The resulting average x ray pulse energy is $2.9\pm0.8$~mJ at the exit of amplifier section for the simulation presented below.

The four-bounce structure used in the stretcher cannot generate positive temporal chirps.
To generate a positive chirp, we adopt the Martinez stretcher \cite{martinez19873000} scheme to the hard x ray wavelengths.
The layout of the compressor is shown in Figure~\ref{fig:layout} (c). 
Crystals $C5$ to $C8$ are arranged mirror-symmetrically with respect to two focusing lenses, $L2$ and $L3$, in between.
To match the bandwidth, $C5$ to $C8$ also use silicon $(111)$ asymmetric Bragg reflections.
The asymmetry angle of $C5$ and $C8$ is $\alpha_1$.
The asymmetry angle of $C6$ and $C7$ is $\alpha_2$ and $\alpha_1 \neq \alpha_2$.
Both focusing lenses have the same focal length, $f$, and form a telescope with a magnification factor of 1.
Because $\alpha_1\neq\alpha_2$, the crystal pair $C5$-$C6$ leads to a net angular dispersion, which is compensated by $C7$-$C8$.
Between $C6$ and $C7$, the telescope converts this angular dispersion into spectral-temporal chirp.
Assume that the path length between the beam footprints within crystal pairs $C5$-$C6$ and $C7$-$C8$ is $l_1$, the distance between crystal $C6$ or $C7$ and their adjacent focusing lens is negligible compared with the focal length $f$.
Then, the energy dependence of the path length in the compressor can be attributed to primarily two components:
\begin{multline}
    \delta(L) \approx
    -8l_1\frac{\sin^2{\alpha_1}\sin^2{\theta}}{\sin^2{\left(\alpha_1+\theta\right)}}\times \frac{\delta E}{E_0} \\
    +8f\frac{\sin^2{\left(\alpha_1-\alpha_2\right)}\sin^4{\theta}}{\sin^2{\left(\alpha_1+\theta\right)}\sin^2{\left(\alpha_2-\theta\right)}}\times\frac{\delta E}{E_0} .
    \label{eq:compressorPL}
\end{multline}
The first term, resembling the energy dependence of the stretcher, shows that the propagation within $C5$-$C6$ and $C7$-$C8$ introduces negative temporal chirp.
The second term represents a positive temporal chirp proportional to the focal length $f$.
By choosing a proper set of $f$, $\alpha_1$, and $\alpha_2$, a net positive temporal chirp can be generated.
The derivation of equation (\ref{eq:compressorPL}) is summarized in Appendix~\ref{appendix:RayTracingCompressor}.

\begin{figure}[bht!]
    \centering
    \includegraphics[width=0.45\textwidth]{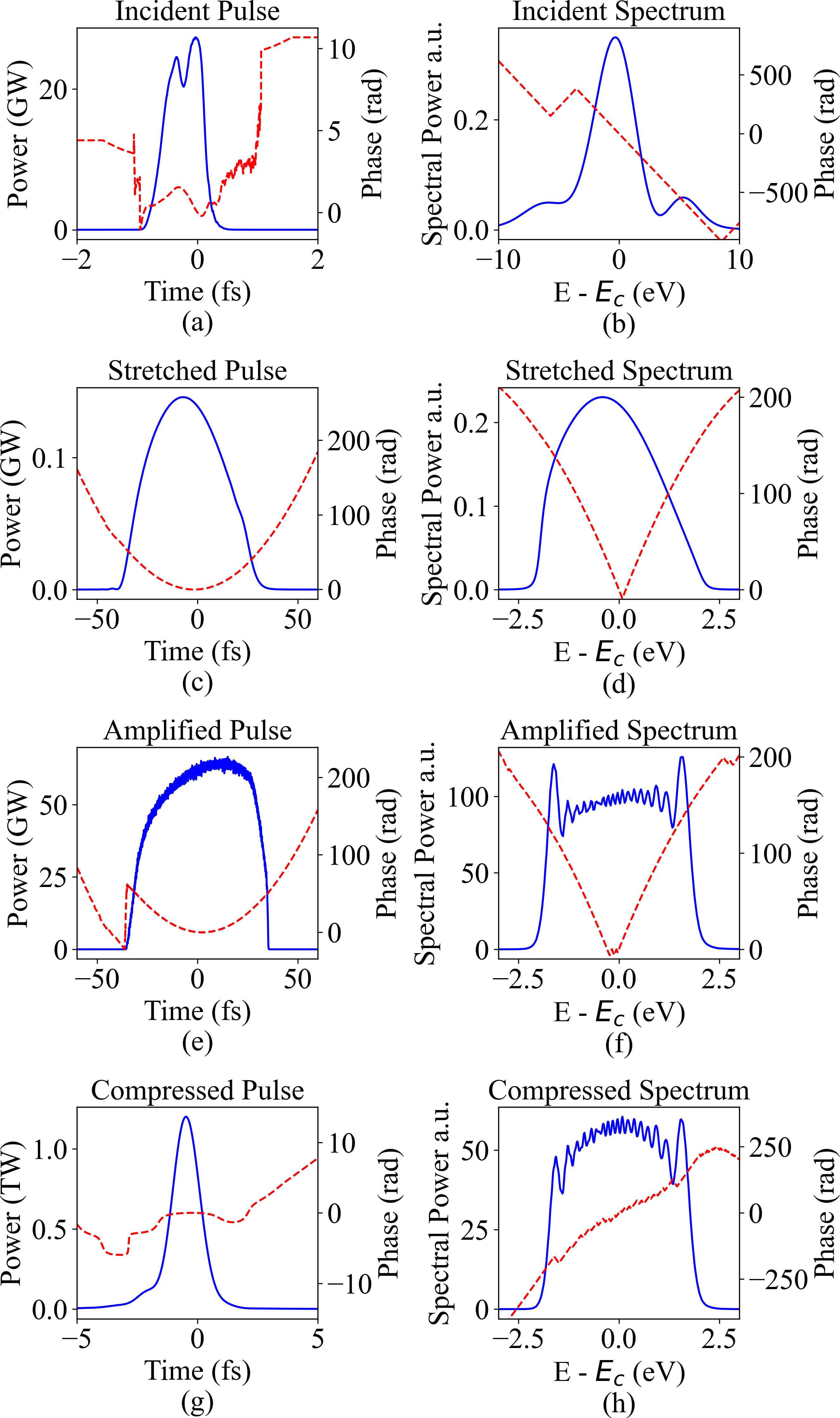}
    \caption{The subplots (a), (c), (e) and (g) show the power profile (blue solid line) and the phase (red dashed line) on-axis.
    The (b), (d), (f) and (h) show the spectrum profile (blue solid line) and the phase (red dashed line) on-axis.
    }
    \label{fig:best_pulse}
\end{figure}

The compressor parameters are determined by the balance between the bandwidth and reflectivity of Bragg reflections with different asymmetry angles.
Bandwidth increase from the amplification process also needs to be considered.
To maximize the final peak power, we surveyed the asymmetry angles $\alpha_2$ of $C6$ and $C7$ from $10\degree$ to $10.7\degree$ with a step size of $0.1\degree$ in the numerical simulation.
In each step, the focal length $f$ is optimized with increments of $0.1~\text{m}$ to obtain the highest peak power.
In this process, we choose $\alpha_1 = \alpha_2+0.2\degree$ to generate a non-zero angular dispersion with $C5$-$C6$ for the telescope to generate the positive temporal chirp.
This angle difference is chosen such that $f$ is neither too short such that the lens loss becomes significant, nor too long such that the whole setup cannot reside within the existing LCLS x ray transport tunnel. 
In the simulation, the transmission function of the two lens are calculated assuming a parabola shape function with the actual complex refractive index of the Beryllium.
This optimization step yielded $\alpha_1=10.7\degree$, $\alpha_2=10.5\degree$, and $f=31.5~\text{m}$.

The angular dispersion introduced by $C5$-$C6$ varies slightly for different incident angles.
This will reduce the compression ratio. 
To mitigate this effect, the amplified pulse is collimated before entering the compressor.
Specifically, after exiting the amplifier section, the amplified pulse propagates $280~\text{m}$ in free space, and passes through a focusing lens, $L1$, with a focal length of $315~\text{m}$.
After the collimation, at the position of $C5$ of the compressor, the averaged FWHM angular divergence of the pulse is $0.1\pm0.04~\mu \text{rad}$, and the average FWHM pulse size is $444.7\pm52.4~\mu \text{m}$, a beam condition where existing silicon monochrometers have operated under similar beam energy ($3$-$4$~mJ) without observable performance degradation.
Figure~\ref{fig:best_pulse} shows the temporal and spectral evolution of one typical simulation using this configuration.
The seed pulse had a FWHM pulse duration of $0.7~\text{fs}$. 
The stretcher extended the pulse to $47.5~\text{fs}$ in FWHM with an energy efficiency of $59\%$ within the bandwidth of 9.498-to-9.502~keV, and $35\%$ over the whole spectrum.
After the amplification, the FWHM pulse duration increased to $62.3~\text{fs}$. 
Finally, after the compression, the FWHM pulse duration was reduced to $1.3~\text{fs}$, with an energy efficiency of $52\%$ through the compressor.
The peak power reached $1.2~\text{TW}$.

\begin{figure}[bht!]
    \centering
    \includegraphics[width=0.45\textwidth]{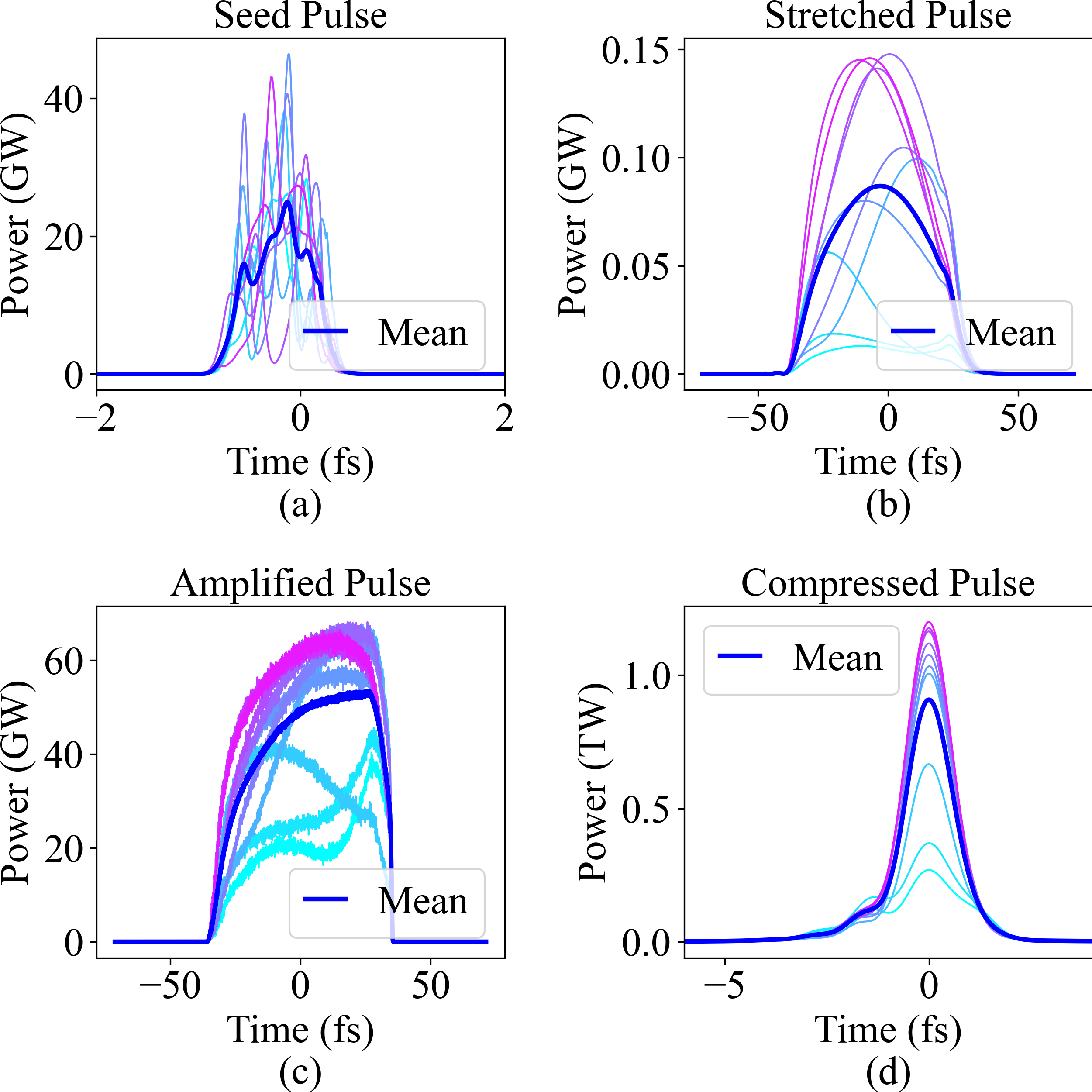}
    \caption{
    Power profiles of 10 simulated pulses at 4 different stages indicated by their titles.
    Thins lines with different colors (except blue) represent each single pulses.
    The thick blue line represent the averaged beam profile of the 10 pulses.
    }
    \label{fig:statistics}
\end{figure}

The statistical distribution of 10 independent simulations are shown in Figure~\ref{fig:statistics}.
The incident FWHM pulse duration was $0.6 \pm 0.2~\text{fs}$ with a median value of $0.6~\text{fs}$.
The stretched pulse duration was $46.9\pm9.2~\text{fs}$ with a median value of $47.8~\text{fs}$, which resulted in a stretching ratio of $79.1$.
The total energy efficiency of the stretcher was $24\% \pm 13\%$ with a median value of $24\%$ over the whole spectrum.
At the exit of the amplifier section, the averaged pulse duration was $56.8\pm 5.7~\text{fs}$ with a median value of $59.7~\text{fs}$.
After the compression, the pulse duration was reduced to $1.4\pm 0.2~\text{fs}$ with a median value of $1.3~\text{fs}$, which corresponded to a compression ratio of $40.6$.
The total energy efficiency of the compressor was $51\% \pm 2\%$ with a median value of $52\%$.
Out of the $10$ pulses, $7$ of them had peak power greater than $1~\text{TW}$, and peak brightness exceeding 4$\times 10^{35}$ s$^{-1}$~mm$^{-2}$~mrad$^{-2}~0.1\%$~bandwidth$^{-1}$.
Besides, as shown in Figure~\ref{fig:statistics} (d), all pulses except the lowest two pulses feature a single main peak without any pre-pulses, which can be easily removed from the dataset by a simple pulse energy filter.
This is of great importance for strong field physics studies, since signals from weak pre-pulses can pose significant challenges in accurate data interpretation. 

To summarize, detailed feasibility study showed that by extending chirped pulse amplification to hard x ray wavelengths using crystal optics based pulse stretcher and compressors, near-transform-limited TW-fs hard x ray pulses can be consistently generated, using electron beam parameters well within the reach of existing free electron laser facilities.
The resulting high peak power and brightness will greatly enhance the performance of many current x ray FEL experiments such as nonlinear x-ray spectroscopy based on various emerging nonlinear x-ray optics phenomena \cite{PhysRevLett.120.223902, PhysRevLett.120.023901, PhysRevLett.127.096801, berger2021extreme, PhysRevLett.127.237402}. This opens up new opportunities to track, understand, and control electronic processes in molecules \cite{tanaka2003ultrafast, tanaka2004simulation, pandey2006simulation, campbell2004simulation, PhysRevB.69.155430}.
The relatively narrow bandwidth of x ray pulses and the resulting clean temporal structure of the x-ray pulses will greatly facilitate the quantitative data interpretation of higher order x ray nonlinear phenomena, a prerequisite for developing future applications.

The proposed optical layout is flexible in optics parameters, can be tailored for different photon energies, and optimized for different existing x ray FEL facility infrastructures.
Further advancement in electron beam parameters, utilization of higher compression ratio, availability of additional undulator length, will enable another leap in peak power and brightness.
The proposed approach is also another example of a concerted-action between electron beam optics and x-ray optics for advancing x ray FEL performance.
One can expect that, in the near future, coherent interplay between crystal optics and free-electron-laser can further enhance our capability to control the lasing dynamics of high brightness electron beams.

The authors thank D. Reis and G. Marcus for helpful discussions.
This work is supported by the U.S. Department of Energy, Office of Science, Office of Basic Energy Sciences under Contract No. DE-AC02-76SF00515.


\begin{thebibliography}{43}%
\makeatletter
\providecommand \@ifxundefined [1]{%
 \@ifx{#1\undefined}
}%
\providecommand \@ifnum [1]{%
 \ifnum #1\expandafter \@firstoftwo
 \else \expandafter \@secondoftwo
 \fi
}%
\providecommand \@ifx [1]{%
 \ifx #1\expandafter \@firstoftwo
 \else \expandafter \@secondoftwo
 \fi
}%
\providecommand \natexlab [1]{#1}%
\providecommand \enquote  [1]{``#1''}%
\providecommand \bibnamefont  [1]{#1}%
\providecommand \bibfnamefont [1]{#1}%
\providecommand \citenamefont [1]{#1}%
\providecommand \href@noop [0]{\@secondoftwo}%
\providecommand \href [0]{\begingroup \@sanitize@url \@href}%
\providecommand \@href[1]{\@@startlink{#1}\@@href}%
\providecommand \@@href[1]{\endgroup#1\@@endlink}%
\providecommand \@sanitize@url [0]{\catcode `\\12\catcode `\$12\catcode
  `\&12\catcode `\#12\catcode `\^12\catcode `\_12\catcode `\%12\relax}%
\providecommand \@@startlink[1]{}%
\providecommand \@@endlink[0]{}%
\providecommand \url  [0]{\begingroup\@sanitize@url \@url }%
\providecommand \@url [1]{\endgroup\@href {#1}{\urlprefix }}%
\providecommand \urlprefix  [0]{URL }%
\providecommand \Eprint [0]{\href }%
\providecommand \doibase [0]{https://doi.org/}%
\providecommand \selectlanguage [0]{\@gobble}%
\providecommand \bibinfo  [0]{\@secondoftwo}%
\providecommand \bibfield  [0]{\@secondoftwo}%
\providecommand \translation [1]{[#1]}%
\providecommand \BibitemOpen [0]{}%
\providecommand \bibitemStop [0]{}%
\providecommand \bibitemNoStop [0]{.\EOS\space}%
\providecommand \EOS [0]{\spacefactor3000\relax}%
\providecommand \BibitemShut  [1]{\csname bibitem#1\endcsname}%
\let\auto@bib@innerbib\@empty
\bibitem [{\citenamefont {Fuchs}\ \emph {et~al.}(2015)\citenamefont {Fuchs},
  \citenamefont {Trigo}, \citenamefont {Chen}, \citenamefont {Ghimire},
  \citenamefont {Shwartz}, \citenamefont {Kozina}, \citenamefont {Jiang},
  \citenamefont {Henighan}, \citenamefont {Bray}, \citenamefont {Ndabashimiye}
  \emph {et~al.}}]{fuchs2015anomalous}%
  \BibitemOpen
  \bibfield  {author} {\bibinfo {author} {\bibfnamefont {M.}~\bibnamefont
  {Fuchs}}, \bibinfo {author} {\bibfnamefont {M.}~\bibnamefont {Trigo}},
  \bibinfo {author} {\bibfnamefont {J.}~\bibnamefont {Chen}}, \bibinfo {author}
  {\bibfnamefont {S.}~\bibnamefont {Ghimire}}, \bibinfo {author} {\bibfnamefont
  {S.}~\bibnamefont {Shwartz}}, \bibinfo {author} {\bibfnamefont
  {M.}~\bibnamefont {Kozina}}, \bibinfo {author} {\bibfnamefont
  {M.}~\bibnamefont {Jiang}}, \bibinfo {author} {\bibfnamefont
  {T.}~\bibnamefont {Henighan}}, \bibinfo {author} {\bibfnamefont
  {C.}~\bibnamefont {Bray}}, \bibinfo {author} {\bibfnamefont {G.}~\bibnamefont
  {Ndabashimiye}}, \emph {et~al.},\ }\bibfield  {title} {\bibinfo {title}
  {Anomalous nonlinear x-ray compton scattering},\ }\href@noop {} {\bibfield
  {journal} {\bibinfo  {journal} {Nature Physics}\ }\textbf {\bibinfo {volume}
  {11}},\ \bibinfo {pages} {964} (\bibinfo {year} {2015})}\BibitemShut
  {NoStop}%
\bibitem [{\citenamefont {Krebs}\ \emph {et~al.}(2019)\citenamefont {Krebs},
  \citenamefont {Reis},\ and\ \citenamefont {Santra}}]{krebs2019time}%
  \BibitemOpen
  \bibfield  {author} {\bibinfo {author} {\bibfnamefont {D.}~\bibnamefont
  {Krebs}}, \bibinfo {author} {\bibfnamefont {D.~A.}\ \bibnamefont {Reis}},\
  and\ \bibinfo {author} {\bibfnamefont {R.}~\bibnamefont {Santra}},\
  }\bibfield  {title} {\bibinfo {title} {Time-dependent qed approach to x-ray
  nonlinear compton scattering},\ }\href@noop {} {\bibfield  {journal}
  {\bibinfo  {journal} {Physical Review A}\ }\textbf {\bibinfo {volume} {99}},\
  \bibinfo {pages} {022120} (\bibinfo {year} {2019})}\BibitemShut {NoStop}%
\bibitem [{\citenamefont {Shwartz}\ \emph {et~al.}(2014)\citenamefont
  {Shwartz}, \citenamefont {Fuchs}, \citenamefont {Hastings}, \citenamefont
  {Inubushi}, \citenamefont {Ishikawa}, \citenamefont {Katayama}, \citenamefont
  {Reis}, \citenamefont {Sato}, \citenamefont {Tono}, \citenamefont {Yabashi},
  \citenamefont {Yudovich},\ and\ \citenamefont
  {Harris}}]{PhysRevLett.112.163901}%
  \BibitemOpen
  \bibfield  {author} {\bibinfo {author} {\bibfnamefont {S.}~\bibnamefont
  {Shwartz}}, \bibinfo {author} {\bibfnamefont {M.}~\bibnamefont {Fuchs}},
  \bibinfo {author} {\bibfnamefont {J.~B.}\ \bibnamefont {Hastings}}, \bibinfo
  {author} {\bibfnamefont {Y.}~\bibnamefont {Inubushi}}, \bibinfo {author}
  {\bibfnamefont {T.}~\bibnamefont {Ishikawa}}, \bibinfo {author}
  {\bibfnamefont {T.}~\bibnamefont {Katayama}}, \bibinfo {author}
  {\bibfnamefont {D.~A.}\ \bibnamefont {Reis}}, \bibinfo {author}
  {\bibfnamefont {T.}~\bibnamefont {Sato}}, \bibinfo {author} {\bibfnamefont
  {K.}~\bibnamefont {Tono}}, \bibinfo {author} {\bibfnamefont {M.}~\bibnamefont
  {Yabashi}}, \bibinfo {author} {\bibfnamefont {S.}~\bibnamefont {Yudovich}},\
  and\ \bibinfo {author} {\bibfnamefont {S.~E.}\ \bibnamefont {Harris}},\
  }\bibfield  {title} {\bibinfo {title} {X-ray second harmonic generation},\
  }\href {https://doi.org/10.1103/PhysRevLett.112.163901} {\bibfield  {journal}
  {\bibinfo  {journal} {Phys. Rev. Lett.}\ }\textbf {\bibinfo {volume} {112}},\
  \bibinfo {pages} {163901} (\bibinfo {year} {2014})}\BibitemShut {NoStop}%
\bibitem [{\citenamefont {Yamamoto}\ \emph {et~al.}(2018)\citenamefont
  {Yamamoto}, \citenamefont {Omi}, \citenamefont {Akai}, \citenamefont
  {Kubota}, \citenamefont {Takahashi}, \citenamefont {Suzuki}, \citenamefont
  {Hirata}, \citenamefont {Yamamoto}, \citenamefont {Yukawa}, \citenamefont
  {Horiba}, \citenamefont {Yumoto}, \citenamefont {Koyama}, \citenamefont
  {Ohashi}, \citenamefont {Owada}, \citenamefont {Tono}, \citenamefont
  {Yabashi}, \citenamefont {Shigemasa}, \citenamefont {Yamamoto}, \citenamefont
  {Kotsugi}, \citenamefont {Wadati}, \citenamefont {Kumigashira}, \citenamefont
  {Arima}, \citenamefont {Shin},\ and\ \citenamefont
  {Matsuda}}]{PhysRevLett.120.223902}%
  \BibitemOpen
  \bibfield  {author} {\bibinfo {author} {\bibfnamefont {S.}~\bibnamefont
  {Yamamoto}}, \bibinfo {author} {\bibfnamefont {T.}~\bibnamefont {Omi}},
  \bibinfo {author} {\bibfnamefont {H.}~\bibnamefont {Akai}}, \bibinfo {author}
  {\bibfnamefont {Y.}~\bibnamefont {Kubota}}, \bibinfo {author} {\bibfnamefont
  {Y.}~\bibnamefont {Takahashi}}, \bibinfo {author} {\bibfnamefont
  {Y.}~\bibnamefont {Suzuki}}, \bibinfo {author} {\bibfnamefont
  {Y.}~\bibnamefont {Hirata}}, \bibinfo {author} {\bibfnamefont
  {K.}~\bibnamefont {Yamamoto}}, \bibinfo {author} {\bibfnamefont
  {R.}~\bibnamefont {Yukawa}}, \bibinfo {author} {\bibfnamefont
  {K.}~\bibnamefont {Horiba}}, \bibinfo {author} {\bibfnamefont
  {H.}~\bibnamefont {Yumoto}}, \bibinfo {author} {\bibfnamefont
  {T.}~\bibnamefont {Koyama}}, \bibinfo {author} {\bibfnamefont
  {H.}~\bibnamefont {Ohashi}}, \bibinfo {author} {\bibfnamefont
  {S.}~\bibnamefont {Owada}}, \bibinfo {author} {\bibfnamefont
  {K.}~\bibnamefont {Tono}}, \bibinfo {author} {\bibfnamefont {M.}~\bibnamefont
  {Yabashi}}, \bibinfo {author} {\bibfnamefont {E.}~\bibnamefont {Shigemasa}},
  \bibinfo {author} {\bibfnamefont {S.}~\bibnamefont {Yamamoto}}, \bibinfo
  {author} {\bibfnamefont {M.}~\bibnamefont {Kotsugi}}, \bibinfo {author}
  {\bibfnamefont {H.}~\bibnamefont {Wadati}}, \bibinfo {author} {\bibfnamefont
  {H.}~\bibnamefont {Kumigashira}}, \bibinfo {author} {\bibfnamefont
  {T.}~\bibnamefont {Arima}}, \bibinfo {author} {\bibfnamefont
  {S.}~\bibnamefont {Shin}},\ and\ \bibinfo {author} {\bibfnamefont
  {I.}~\bibnamefont {Matsuda}},\ }\bibfield  {title} {\bibinfo {title} {Element
  selectivity in second-harmonic generation of ${\mathrm{gafeo}}_{3}$ by a
  soft-x-ray free-electron laser},\ }\href
  {https://doi.org/10.1103/PhysRevLett.120.223902} {\bibfield  {journal}
  {\bibinfo  {journal} {Phys. Rev. Lett.}\ }\textbf {\bibinfo {volume} {120}},\
  \bibinfo {pages} {223902} (\bibinfo {year} {2018})}\BibitemShut {NoStop}%
\bibitem [{\citenamefont {Lam}\ \emph {et~al.}(2018)\citenamefont {Lam},
  \citenamefont {Raj}, \citenamefont {Pascal}, \citenamefont {Pemmaraju},
  \citenamefont {Foglia}, \citenamefont {Simoncig}, \citenamefont {Fabris},
  \citenamefont {Miotti}, \citenamefont {Hull}, \citenamefont {Rizzuto},
  \citenamefont {Smith}, \citenamefont {Mincigrucci}, \citenamefont
  {Masciovecchio}, \citenamefont {Gessini}, \citenamefont {Allaria},
  \citenamefont {De~Ninno}, \citenamefont {Diviacco}, \citenamefont {Roussel},
  \citenamefont {Spampinati}, \citenamefont {Penco}, \citenamefont {Di~Mitri},
  \citenamefont {Trov\`o}, \citenamefont {Danailov}, \citenamefont
  {Christensen}, \citenamefont {Sokaras}, \citenamefont {Weng}, \citenamefont
  {Coreno}, \citenamefont {Poletto}, \citenamefont {Drisdell}, \citenamefont
  {Prendergast}, \citenamefont {Giannessi}, \citenamefont {Principi},
  \citenamefont {Nordlund}, \citenamefont {Saykally},\ and\ \citenamefont
  {Schwartz}}]{PhysRevLett.120.023901}%
  \BibitemOpen
  \bibfield  {author} {\bibinfo {author} {\bibfnamefont {R.~K.}\ \bibnamefont
  {Lam}}, \bibinfo {author} {\bibfnamefont {S.~L.}\ \bibnamefont {Raj}},
  \bibinfo {author} {\bibfnamefont {T.~A.}\ \bibnamefont {Pascal}}, \bibinfo
  {author} {\bibfnamefont {C.~D.}\ \bibnamefont {Pemmaraju}}, \bibinfo {author}
  {\bibfnamefont {L.}~\bibnamefont {Foglia}}, \bibinfo {author} {\bibfnamefont
  {A.}~\bibnamefont {Simoncig}}, \bibinfo {author} {\bibfnamefont
  {N.}~\bibnamefont {Fabris}}, \bibinfo {author} {\bibfnamefont
  {P.}~\bibnamefont {Miotti}}, \bibinfo {author} {\bibfnamefont {C.~J.}\
  \bibnamefont {Hull}}, \bibinfo {author} {\bibfnamefont {A.~M.}\ \bibnamefont
  {Rizzuto}}, \bibinfo {author} {\bibfnamefont {J.~W.}\ \bibnamefont {Smith}},
  \bibinfo {author} {\bibfnamefont {R.}~\bibnamefont {Mincigrucci}}, \bibinfo
  {author} {\bibfnamefont {C.}~\bibnamefont {Masciovecchio}}, \bibinfo {author}
  {\bibfnamefont {A.}~\bibnamefont {Gessini}}, \bibinfo {author} {\bibfnamefont
  {E.}~\bibnamefont {Allaria}}, \bibinfo {author} {\bibfnamefont
  {G.}~\bibnamefont {De~Ninno}}, \bibinfo {author} {\bibfnamefont
  {B.}~\bibnamefont {Diviacco}}, \bibinfo {author} {\bibfnamefont
  {E.}~\bibnamefont {Roussel}}, \bibinfo {author} {\bibfnamefont
  {S.}~\bibnamefont {Spampinati}}, \bibinfo {author} {\bibfnamefont
  {G.}~\bibnamefont {Penco}}, \bibinfo {author} {\bibfnamefont
  {S.}~\bibnamefont {Di~Mitri}}, \bibinfo {author} {\bibfnamefont
  {M.}~\bibnamefont {Trov\`o}}, \bibinfo {author} {\bibfnamefont
  {M.}~\bibnamefont {Danailov}}, \bibinfo {author} {\bibfnamefont {S.~T.}\
  \bibnamefont {Christensen}}, \bibinfo {author} {\bibfnamefont
  {D.}~\bibnamefont {Sokaras}}, \bibinfo {author} {\bibfnamefont {T.-C.}\
  \bibnamefont {Weng}}, \bibinfo {author} {\bibfnamefont {M.}~\bibnamefont
  {Coreno}}, \bibinfo {author} {\bibfnamefont {L.}~\bibnamefont {Poletto}},
  \bibinfo {author} {\bibfnamefont {W.~S.}\ \bibnamefont {Drisdell}}, \bibinfo
  {author} {\bibfnamefont {D.}~\bibnamefont {Prendergast}}, \bibinfo {author}
  {\bibfnamefont {L.}~\bibnamefont {Giannessi}}, \bibinfo {author}
  {\bibfnamefont {E.}~\bibnamefont {Principi}}, \bibinfo {author}
  {\bibfnamefont {D.}~\bibnamefont {Nordlund}}, \bibinfo {author}
  {\bibfnamefont {R.~J.}\ \bibnamefont {Saykally}},\ and\ \bibinfo {author}
  {\bibfnamefont {C.~P.}\ \bibnamefont {Schwartz}},\ }\bibfield  {title}
  {\bibinfo {title} {Soft x-ray second harmonic generation as an interfacial
  probe},\ }\href {https://doi.org/10.1103/PhysRevLett.120.023901} {\bibfield
  {journal} {\bibinfo  {journal} {Phys. Rev. Lett.}\ }\textbf {\bibinfo
  {volume} {120}},\ \bibinfo {pages} {023901} (\bibinfo {year}
  {2018})}\BibitemShut {NoStop}%
\bibitem [{\citenamefont {Schori}\ \emph {et~al.}(2017)\citenamefont {Schori},
  \citenamefont {B\"omer}, \citenamefont {Borodin}, \citenamefont {Collins},
  \citenamefont {Detlefs}, \citenamefont {Moretti~Sala}, \citenamefont
  {Yudovich},\ and\ \citenamefont {Shwartz}}]{schori2017parametric}%
  \BibitemOpen
  \bibfield  {author} {\bibinfo {author} {\bibfnamefont {A.}~\bibnamefont
  {Schori}}, \bibinfo {author} {\bibfnamefont {C.}~\bibnamefont {B\"omer}},
  \bibinfo {author} {\bibfnamefont {D.}~\bibnamefont {Borodin}}, \bibinfo
  {author} {\bibfnamefont {S.~P.}\ \bibnamefont {Collins}}, \bibinfo {author}
  {\bibfnamefont {B.}~\bibnamefont {Detlefs}}, \bibinfo {author} {\bibfnamefont
  {M.}~\bibnamefont {Moretti~Sala}}, \bibinfo {author} {\bibfnamefont
  {S.}~\bibnamefont {Yudovich}},\ and\ \bibinfo {author} {\bibfnamefont
  {S.}~\bibnamefont {Shwartz}},\ }\bibfield  {title} {\bibinfo {title}
  {Parametric down-conversion of x rays into the optical regime},\ }\href
  {https://doi.org/10.1103/PhysRevLett.119.253902} {\bibfield  {journal}
  {\bibinfo  {journal} {Phys. Rev. Lett.}\ }\textbf {\bibinfo {volume} {119}},\
  \bibinfo {pages} {253902} (\bibinfo {year} {2017})}\BibitemShut {NoStop}%
\bibitem [{\citenamefont {Sofer}\ \emph {et~al.}(2019)\citenamefont {Sofer},
  \citenamefont {Sefi}, \citenamefont {Strizhevsky}, \citenamefont {Aknin},
  \citenamefont {Collins}, \citenamefont {Nisbet}, \citenamefont {Detlefs},
  \citenamefont {Sahle},\ and\ \citenamefont {Shwartz}}]{sofer2019observation}%
  \BibitemOpen
  \bibfield  {author} {\bibinfo {author} {\bibfnamefont {S.}~\bibnamefont
  {Sofer}}, \bibinfo {author} {\bibfnamefont {O.}~\bibnamefont {Sefi}},
  \bibinfo {author} {\bibfnamefont {E.}~\bibnamefont {Strizhevsky}}, \bibinfo
  {author} {\bibfnamefont {H.}~\bibnamefont {Aknin}}, \bibinfo {author}
  {\bibfnamefont {S.}~\bibnamefont {Collins}}, \bibinfo {author} {\bibfnamefont
  {G.}~\bibnamefont {Nisbet}}, \bibinfo {author} {\bibfnamefont
  {B.}~\bibnamefont {Detlefs}}, \bibinfo {author} {\bibfnamefont {C.~J.}\
  \bibnamefont {Sahle}},\ and\ \bibinfo {author} {\bibfnamefont
  {S.}~\bibnamefont {Shwartz}},\ }\bibfield  {title} {\bibinfo {title}
  {Observation of strong nonlinear interactions in parametric down-conversion
  of x-rays into ultraviolet radiation},\ }\href@noop {} {\bibfield  {journal}
  {\bibinfo  {journal} {Nature communications}\ }\textbf {\bibinfo {volume}
  {10}},\ \bibinfo {pages} {1} (\bibinfo {year} {2019})}\BibitemShut {NoStop}%
\bibitem [{\citenamefont {Krebs}\ and\ \citenamefont
  {Rohringer}(2021)}]{krebs2021theory}%
  \BibitemOpen
  \bibfield  {author} {\bibinfo {author} {\bibfnamefont {D.}~\bibnamefont
  {Krebs}}\ and\ \bibinfo {author} {\bibfnamefont {N.}~\bibnamefont
  {Rohringer}},\ }\bibfield  {title} {\bibinfo {title} {Theory of parametric
  x-ray optical wavemixing processes},\ }\href@noop {} {\bibfield  {journal}
  {\bibinfo  {journal} {arXiv preprint arXiv:2104.05838}\ } (\bibinfo {year}
  {2021})}\BibitemShut {NoStop}%
\bibitem [{\citenamefont {Kroll}\ \emph {et~al.}(2018)\citenamefont {Kroll},
  \citenamefont {Weninger}, \citenamefont {Alonso-Mori}, \citenamefont
  {Sokaras}, \citenamefont {Zhu}, \citenamefont {Mercadier}, \citenamefont
  {Majety}, \citenamefont {Marinelli}, \citenamefont {Lutman}, \citenamefont
  {Guetg}, \citenamefont {Decker}, \citenamefont {Boutet}, \citenamefont
  {Aquila}, \citenamefont {Koglin}, \citenamefont {Koralek}, \citenamefont
  {DePonte}, \citenamefont {Kern}, \citenamefont {Fuller}, \citenamefont
  {Pastor}, \citenamefont {Fransson}, \citenamefont {Zhang}, \citenamefont
  {Yano}, \citenamefont {Yachandra}, \citenamefont {Rohringer},\ and\
  \citenamefont {Bergmann}}]{kroll2018stimulated}%
  \BibitemOpen
  \bibfield  {author} {\bibinfo {author} {\bibfnamefont {T.}~\bibnamefont
  {Kroll}}, \bibinfo {author} {\bibfnamefont {C.}~\bibnamefont {Weninger}},
  \bibinfo {author} {\bibfnamefont {R.}~\bibnamefont {Alonso-Mori}}, \bibinfo
  {author} {\bibfnamefont {D.}~\bibnamefont {Sokaras}}, \bibinfo {author}
  {\bibfnamefont {D.}~\bibnamefont {Zhu}}, \bibinfo {author} {\bibfnamefont
  {L.}~\bibnamefont {Mercadier}}, \bibinfo {author} {\bibfnamefont {V.~P.}\
  \bibnamefont {Majety}}, \bibinfo {author} {\bibfnamefont {A.}~\bibnamefont
  {Marinelli}}, \bibinfo {author} {\bibfnamefont {A.}~\bibnamefont {Lutman}},
  \bibinfo {author} {\bibfnamefont {M.~W.}\ \bibnamefont {Guetg}}, \bibinfo
  {author} {\bibfnamefont {F.-J.}\ \bibnamefont {Decker}}, \bibinfo {author}
  {\bibfnamefont {S.}~\bibnamefont {Boutet}}, \bibinfo {author} {\bibfnamefont
  {A.}~\bibnamefont {Aquila}}, \bibinfo {author} {\bibfnamefont
  {J.}~\bibnamefont {Koglin}}, \bibinfo {author} {\bibfnamefont
  {J.}~\bibnamefont {Koralek}}, \bibinfo {author} {\bibfnamefont {D.~P.}\
  \bibnamefont {DePonte}}, \bibinfo {author} {\bibfnamefont {J.}~\bibnamefont
  {Kern}}, \bibinfo {author} {\bibfnamefont {F.~D.}\ \bibnamefont {Fuller}},
  \bibinfo {author} {\bibfnamefont {E.}~\bibnamefont {Pastor}}, \bibinfo
  {author} {\bibfnamefont {T.}~\bibnamefont {Fransson}}, \bibinfo {author}
  {\bibfnamefont {Y.}~\bibnamefont {Zhang}}, \bibinfo {author} {\bibfnamefont
  {J.}~\bibnamefont {Yano}}, \bibinfo {author} {\bibfnamefont {V.~K.}\
  \bibnamefont {Yachandra}}, \bibinfo {author} {\bibfnamefont {N.}~\bibnamefont
  {Rohringer}},\ and\ \bibinfo {author} {\bibfnamefont {U.}~\bibnamefont
  {Bergmann}},\ }\bibfield  {title} {\bibinfo {title} {Stimulated x-ray
  emission spectroscopy in transition metal complexes},\ }\href
  {https://doi.org/10.1103/PhysRevLett.120.133203} {\bibfield  {journal}
  {\bibinfo  {journal} {Phys. Rev. Lett.}\ }\textbf {\bibinfo {volume} {120}},\
  \bibinfo {pages} {133203} (\bibinfo {year} {2018})}\BibitemShut {NoStop}%
\bibitem [{\citenamefont {Glover}\ \emph {et~al.}(2012)\citenamefont {Glover},
  \citenamefont {Fritz}, \citenamefont {Cammarata}, \citenamefont {Allison},
  \citenamefont {Coh}, \citenamefont {Feldkamp}, \citenamefont {Lemke},
  \citenamefont {Zhu}, \citenamefont {Feng}, \citenamefont {Coffee} \emph
  {et~al.}}]{glover2012x}%
  \BibitemOpen
  \bibfield  {author} {\bibinfo {author} {\bibfnamefont {T.}~\bibnamefont
  {Glover}}, \bibinfo {author} {\bibfnamefont {D.}~\bibnamefont {Fritz}},
  \bibinfo {author} {\bibfnamefont {M.}~\bibnamefont {Cammarata}}, \bibinfo
  {author} {\bibfnamefont {T.}~\bibnamefont {Allison}}, \bibinfo {author}
  {\bibfnamefont {S.}~\bibnamefont {Coh}}, \bibinfo {author} {\bibfnamefont
  {J.}~\bibnamefont {Feldkamp}}, \bibinfo {author} {\bibfnamefont
  {H.}~\bibnamefont {Lemke}}, \bibinfo {author} {\bibfnamefont
  {D.}~\bibnamefont {Zhu}}, \bibinfo {author} {\bibfnamefont {Y.}~\bibnamefont
  {Feng}}, \bibinfo {author} {\bibfnamefont {R.}~\bibnamefont {Coffee}}, \emph
  {et~al.},\ }\bibfield  {title} {\bibinfo {title} {X-ray and optical wave
  mixing},\ }\href@noop {} {\bibfield  {journal} {\bibinfo  {journal} {Nature}\
  }\textbf {\bibinfo {volume} {488}},\ \bibinfo {pages} {603} (\bibinfo {year}
  {2012})}\BibitemShut {NoStop}%
\bibitem [{\citenamefont {Marcus}\ \emph {et~al.}(2014)\citenamefont {Marcus},
  \citenamefont {Penn},\ and\ \citenamefont {Zholents}}]{marcus2014free}%
  \BibitemOpen
  \bibfield  {author} {\bibinfo {author} {\bibfnamefont {G.}~\bibnamefont
  {Marcus}}, \bibinfo {author} {\bibfnamefont {G.}~\bibnamefont {Penn}},\ and\
  \bibinfo {author} {\bibfnamefont {A.~A.}\ \bibnamefont {Zholents}},\
  }\bibfield  {title} {\bibinfo {title} {Free-electron laser design for
  four-wave mixing experiments with soft-x-ray pulses},\ }\href@noop {}
  {\bibfield  {journal} {\bibinfo  {journal} {Physical review letters}\
  }\textbf {\bibinfo {volume} {113}},\ \bibinfo {pages} {024801} (\bibinfo
  {year} {2014})}\BibitemShut {NoStop}%
\bibitem [{\citenamefont {Mukamel}(2005)}]{PhysRevB.72.235110}%
  \BibitemOpen
  \bibfield  {author} {\bibinfo {author} {\bibfnamefont {S.}~\bibnamefont
  {Mukamel}},\ }\bibfield  {title} {\bibinfo {title} {Multiple core-hole
  coherence in x-ray four-wave-mixing spectroscopies},\ }\href
  {https://doi.org/10.1103/PhysRevB.72.235110} {\bibfield  {journal} {\bibinfo
  {journal} {Phys. Rev. B}\ }\textbf {\bibinfo {volume} {72}},\ \bibinfo
  {pages} {235110} (\bibinfo {year} {2005})}\BibitemShut {NoStop}%
\bibitem [{\citenamefont {Saldin}\ \emph {et~al.}(1998)\citenamefont {Saldin},
  \citenamefont {Schneidmiller},\ and\ \citenamefont
  {Yurkov}}]{saldin1998statistical}%
  \BibitemOpen
  \bibfield  {author} {\bibinfo {author} {\bibfnamefont {E.}~\bibnamefont
  {Saldin}}, \bibinfo {author} {\bibfnamefont {E.}~\bibnamefont
  {Schneidmiller}},\ and\ \bibinfo {author} {\bibfnamefont {M.}~\bibnamefont
  {Yurkov}},\ }\bibfield  {title} {\bibinfo {title} {Statistical properties of
  the radiation from sase fel operating in the linear regime},\ }\href@noop {}
  {\bibfield  {journal} {\bibinfo  {journal} {Nuclear Instruments and Methods
  in Physics Research Section A: Accelerators, Spectrometers, Detectors and
  Associated Equipment}\ }\textbf {\bibinfo {volume} {407}},\ \bibinfo {pages}
  {291} (\bibinfo {year} {1998})}\BibitemShut {NoStop}%
\bibitem [{\citenamefont {Guetg}\ \emph {et~al.}(2018)\citenamefont {Guetg},
  \citenamefont {Lutman}, \citenamefont {Ding}, \citenamefont {Maxwell},
  \citenamefont {Decker}, \citenamefont {Bergmann},\ and\ \citenamefont
  {Huang}}]{PhysRevLett.120.014801}%
  \BibitemOpen
  \bibfield  {author} {\bibinfo {author} {\bibfnamefont {M.~W.}\ \bibnamefont
  {Guetg}}, \bibinfo {author} {\bibfnamefont {A.~A.}\ \bibnamefont {Lutman}},
  \bibinfo {author} {\bibfnamefont {Y.}~\bibnamefont {Ding}}, \bibinfo {author}
  {\bibfnamefont {T.~J.}\ \bibnamefont {Maxwell}}, \bibinfo {author}
  {\bibfnamefont {F.-J.}\ \bibnamefont {Decker}}, \bibinfo {author}
  {\bibfnamefont {U.}~\bibnamefont {Bergmann}},\ and\ \bibinfo {author}
  {\bibfnamefont {Z.}~\bibnamefont {Huang}},\ }\bibfield  {title} {\bibinfo
  {title} {Generation of high-power high-intensity short x-ray
  free-electron-laser pulses},\ }\href
  {https://doi.org/10.1103/PhysRevLett.120.014801} {\bibfield  {journal}
  {\bibinfo  {journal} {Phys. Rev. Lett.}\ }\textbf {\bibinfo {volume} {120}},\
  \bibinfo {pages} {014801} (\bibinfo {year} {2018})}\BibitemShut {NoStop}%
\bibitem [{\citenamefont {Zholents}(2005)}]{PhysRevSTAB.8.040701}%
  \BibitemOpen
  \bibfield  {author} {\bibinfo {author} {\bibfnamefont {A.~A.}\ \bibnamefont
  {Zholents}},\ }\bibfield  {title} {\bibinfo {title} {Method of an enhanced
  self-amplified spontaneous emission for x-ray free electron lasers},\ }\href
  {https://doi.org/10.1103/PhysRevSTAB.8.040701} {\bibfield  {journal}
  {\bibinfo  {journal} {Phys. Rev. ST Accel. Beams}\ }\textbf {\bibinfo
  {volume} {8}},\ \bibinfo {pages} {040701} (\bibinfo {year}
  {2005})}\BibitemShut {NoStop}%
\bibitem [{\citenamefont {Kumar}\ \emph {et~al.}(2016)\citenamefont {Kumar},
  \citenamefont {Parc}, \citenamefont {Landsman},\ and\ \citenamefont
  {Kim}}]{kumar2016temporally}%
  \BibitemOpen
  \bibfield  {author} {\bibinfo {author} {\bibfnamefont {S.}~\bibnamefont
  {Kumar}}, \bibinfo {author} {\bibfnamefont {Y.~W.}\ \bibnamefont {Parc}},
  \bibinfo {author} {\bibfnamefont {A.~S.}\ \bibnamefont {Landsman}},\ and\
  \bibinfo {author} {\bibfnamefont {D.~E.}\ \bibnamefont {Kim}},\ }\bibfield
  {title} {\bibinfo {title} {Temporally-coherent terawatt attosecond xfel
  synchronized with a few cycle laser},\ }\href@noop {} {\bibfield  {journal}
  {\bibinfo  {journal} {Scientific Reports}\ }\textbf {\bibinfo {volume} {6}},\
  \bibinfo {pages} {1} (\bibinfo {year} {2016})}\BibitemShut {NoStop}%
\bibitem [{\citenamefont {Shim}\ \emph {et~al.}(2018)\citenamefont {Shim},
  \citenamefont {Parc}, \citenamefont {Kumar}, \citenamefont {Ko},\ and\
  \citenamefont {Kim}}]{shim2018isolated}%
  \BibitemOpen
  \bibfield  {author} {\bibinfo {author} {\bibfnamefont {C.~H.}\ \bibnamefont
  {Shim}}, \bibinfo {author} {\bibfnamefont {Y.~W.}\ \bibnamefont {Parc}},
  \bibinfo {author} {\bibfnamefont {S.}~\bibnamefont {Kumar}}, \bibinfo
  {author} {\bibfnamefont {I.~S.}\ \bibnamefont {Ko}},\ and\ \bibinfo {author}
  {\bibfnamefont {D.~E.}\ \bibnamefont {Kim}},\ }\bibfield  {title} {\bibinfo
  {title} {Isolated terawatt attosecond hard x-ray pulse generated from single
  current spike},\ }\href@noop {} {\bibfield  {journal} {\bibinfo  {journal}
  {Scientific reports}\ }\textbf {\bibinfo {volume} {8}},\ \bibinfo {pages} {1}
  (\bibinfo {year} {2018})}\BibitemShut {NoStop}%
\bibitem [{\citenamefont {Prat}\ and\ \citenamefont
  {Reiche}(2015)}]{prat2015simple}%
  \BibitemOpen
  \bibfield  {author} {\bibinfo {author} {\bibfnamefont {E.}~\bibnamefont
  {Prat}}\ and\ \bibinfo {author} {\bibfnamefont {S.}~\bibnamefont {Reiche}},\
  }\bibfield  {title} {\bibinfo {title} {Simple method to generate
  terawatt-attosecond x-ray free-electron-laser pulses},\ }\href@noop {}
  {\bibfield  {journal} {\bibinfo  {journal} {Physical Review Letters}\
  }\textbf {\bibinfo {volume} {114}},\ \bibinfo {pages} {244801} (\bibinfo
  {year} {2015})}\BibitemShut {NoStop}%
\bibitem [{\citenamefont {Amann}\ \emph {et~al.}(2012)\citenamefont {Amann},
  \citenamefont {Berg}, \citenamefont {Blank}, \citenamefont {Decker},
  \citenamefont {Ding}, \citenamefont {Emma}, \citenamefont {Feng},
  \citenamefont {Frisch}, \citenamefont {Fritz}, \citenamefont {Hastings} \emph
  {et~al.}}]{amann2012demonstration}%
  \BibitemOpen
  \bibfield  {author} {\bibinfo {author} {\bibfnamefont {J.}~\bibnamefont
  {Amann}}, \bibinfo {author} {\bibfnamefont {W.}~\bibnamefont {Berg}},
  \bibinfo {author} {\bibfnamefont {V.}~\bibnamefont {Blank}}, \bibinfo
  {author} {\bibfnamefont {F.-J.}\ \bibnamefont {Decker}}, \bibinfo {author}
  {\bibfnamefont {Y.}~\bibnamefont {Ding}}, \bibinfo {author} {\bibfnamefont
  {P.}~\bibnamefont {Emma}}, \bibinfo {author} {\bibfnamefont {Y.}~\bibnamefont
  {Feng}}, \bibinfo {author} {\bibfnamefont {J.}~\bibnamefont {Frisch}},
  \bibinfo {author} {\bibfnamefont {D.}~\bibnamefont {Fritz}}, \bibinfo
  {author} {\bibfnamefont {J.}~\bibnamefont {Hastings}}, \emph {et~al.},\
  }\bibfield  {title} {\bibinfo {title} {Demonstration of self-seeding in a
  hard-x-ray free-electron laser},\ }\href@noop {} {\bibfield  {journal}
  {\bibinfo  {journal} {Nature photonics}\ }\textbf {\bibinfo {volume} {6}},\
  \bibinfo {pages} {693} (\bibinfo {year} {2012})}\BibitemShut {NoStop}%
\bibitem [{\citenamefont {Inoue}\ \emph {et~al.}(2019)\citenamefont {Inoue},
  \citenamefont {Osaka}, \citenamefont {Hara}, \citenamefont {Tanaka},
  \citenamefont {Inagaki}, \citenamefont {Fukui}, \citenamefont {Goto},
  \citenamefont {Inubushi}, \citenamefont {Kimura}, \citenamefont {Kinjo} \emph
  {et~al.}}]{inoue2019generation}%
  \BibitemOpen
  \bibfield  {author} {\bibinfo {author} {\bibfnamefont {I.}~\bibnamefont
  {Inoue}}, \bibinfo {author} {\bibfnamefont {T.}~\bibnamefont {Osaka}},
  \bibinfo {author} {\bibfnamefont {T.}~\bibnamefont {Hara}}, \bibinfo {author}
  {\bibfnamefont {T.}~\bibnamefont {Tanaka}}, \bibinfo {author} {\bibfnamefont
  {T.}~\bibnamefont {Inagaki}}, \bibinfo {author} {\bibfnamefont
  {T.}~\bibnamefont {Fukui}}, \bibinfo {author} {\bibfnamefont
  {S.}~\bibnamefont {Goto}}, \bibinfo {author} {\bibfnamefont {Y.}~\bibnamefont
  {Inubushi}}, \bibinfo {author} {\bibfnamefont {H.}~\bibnamefont {Kimura}},
  \bibinfo {author} {\bibfnamefont {R.}~\bibnamefont {Kinjo}}, \emph {et~al.},\
  }\bibfield  {title} {\bibinfo {title} {Generation of narrow-band x-ray
  free-electron laser via reflection self-seeding},\ }\href@noop {} {\bibfield
  {journal} {\bibinfo  {journal} {Nature Photonics}\ }\textbf {\bibinfo
  {volume} {13}},\ \bibinfo {pages} {319} (\bibinfo {year} {2019})}\BibitemShut
  {NoStop}%
\bibitem [{\citenamefont {Min}\ \emph {et~al.}(2019)\citenamefont {Min},
  \citenamefont {Nam}, \citenamefont {Yang}, \citenamefont {Kim}, \citenamefont
  {Shim}, \citenamefont {Ko}, \citenamefont {Cho}, \citenamefont {Heo},
  \citenamefont {Oh}, \citenamefont {Suh} \emph {et~al.}}]{min2019hard}%
  \BibitemOpen
  \bibfield  {author} {\bibinfo {author} {\bibfnamefont {C.-K.}\ \bibnamefont
  {Min}}, \bibinfo {author} {\bibfnamefont {I.}~\bibnamefont {Nam}}, \bibinfo
  {author} {\bibfnamefont {H.}~\bibnamefont {Yang}}, \bibinfo {author}
  {\bibfnamefont {G.}~\bibnamefont {Kim}}, \bibinfo {author} {\bibfnamefont
  {C.~H.}\ \bibnamefont {Shim}}, \bibinfo {author} {\bibfnamefont {J.~H.}\
  \bibnamefont {Ko}}, \bibinfo {author} {\bibfnamefont {M.-H.}\ \bibnamefont
  {Cho}}, \bibinfo {author} {\bibfnamefont {H.}~\bibnamefont {Heo}}, \bibinfo
  {author} {\bibfnamefont {B.}~\bibnamefont {Oh}}, \bibinfo {author}
  {\bibfnamefont {Y.~J.}\ \bibnamefont {Suh}}, \emph {et~al.},\ }\bibfield
  {title} {\bibinfo {title} {Hard x-ray self-seeding commissioning at
  pal-xfel},\ }\href@noop {} {\bibfield  {journal} {\bibinfo  {journal}
  {Journal of synchrotron radiation}\ }\textbf {\bibinfo {volume} {26}},\
  \bibinfo {pages} {1101} (\bibinfo {year} {2019})}\BibitemShut {NoStop}%
\bibitem [{\citenamefont {Halavanau}\ \emph {et~al.}(2019)\citenamefont
  {Halavanau}, \citenamefont {Decker}, \citenamefont {Ding}, \citenamefont
  {Emma}, \citenamefont {Huang}, \citenamefont {Krzywinski}, \citenamefont
  {Lutman}, \citenamefont {Marcus}, \citenamefont {Pellegrini},\ and\
  \citenamefont {Zhu}}]{osti_1560969}%
  \BibitemOpen
  \bibfield  {author} {\bibinfo {author} {\bibfnamefont {A.}~\bibnamefont
  {Halavanau}}, \bibinfo {author} {\bibfnamefont {F.~J.}\ \bibnamefont
  {Decker}}, \bibinfo {author} {\bibfnamefont {Y.}~\bibnamefont {Ding}},
  \bibinfo {author} {\bibfnamefont {C.}~\bibnamefont {Emma}}, \bibinfo {author}
  {\bibfnamefont {Z.}~\bibnamefont {Huang}}, \bibinfo {author} {\bibfnamefont
  {J.}~\bibnamefont {Krzywinski}}, \bibinfo {author} {\bibfnamefont {A.~A.}\
  \bibnamefont {Lutman}}, \bibinfo {author} {\bibfnamefont {G.}~\bibnamefont
  {Marcus}}, \bibinfo {author} {\bibfnamefont {C.}~\bibnamefont {Pellegrini}},\
  and\ \bibinfo {author} {\bibfnamefont {D.}~\bibnamefont {Zhu}},\ }\bibfield
  {title} {\bibinfo {title} {Generation of high peak power hard x-rays at
  lcls-ii with double bunch self-seeding}\ }\href
  {https://www.osti.gov/biblio/1560969} {} (\bibinfo {year} {2019})\BibitemShut
  {NoStop}%
\bibitem [{\citenamefont {Strickland}\ and\ \citenamefont
  {Mourou}(1985)}]{strickland1985compression}%
  \BibitemOpen
  \bibfield  {author} {\bibinfo {author} {\bibfnamefont {D.}~\bibnamefont
  {Strickland}}\ and\ \bibinfo {author} {\bibfnamefont {G.}~\bibnamefont
  {Mourou}},\ }\bibfield  {title} {\bibinfo {title} {Compression of amplified
  chirped optical pulses},\ }\href@noop {} {\bibfield  {journal} {\bibinfo
  {journal} {Optics communications}\ }\textbf {\bibinfo {volume} {55}},\
  \bibinfo {pages} {447} (\bibinfo {year} {1985})}\BibitemShut {NoStop}%
\bibitem [{\citenamefont {Papadopoulos}\ \emph {et~al.}(2016)\citenamefont
  {Papadopoulos}, \citenamefont {Zou}, \citenamefont {Le~Blanc}, \citenamefont
  {Ch{\'e}riaux}, \citenamefont {Georges}, \citenamefont {Druon}, \citenamefont
  {Mennerat}, \citenamefont {Ramirez}, \citenamefont {Martin}, \citenamefont
  {Fr{\'e}neaux} \emph {et~al.}}]{papadopoulos2016apollon}%
  \BibitemOpen
  \bibfield  {author} {\bibinfo {author} {\bibfnamefont {D.}~\bibnamefont
  {Papadopoulos}}, \bibinfo {author} {\bibfnamefont {J.}~\bibnamefont {Zou}},
  \bibinfo {author} {\bibfnamefont {C.}~\bibnamefont {Le~Blanc}}, \bibinfo
  {author} {\bibfnamefont {G.}~\bibnamefont {Ch{\'e}riaux}}, \bibinfo {author}
  {\bibfnamefont {P.}~\bibnamefont {Georges}}, \bibinfo {author} {\bibfnamefont
  {F.}~\bibnamefont {Druon}}, \bibinfo {author} {\bibfnamefont
  {G.}~\bibnamefont {Mennerat}}, \bibinfo {author} {\bibfnamefont
  {P.}~\bibnamefont {Ramirez}}, \bibinfo {author} {\bibfnamefont
  {L.}~\bibnamefont {Martin}}, \bibinfo {author} {\bibfnamefont
  {A.}~\bibnamefont {Fr{\'e}neaux}}, \emph {et~al.},\ }\bibfield  {title}
  {\bibinfo {title} {The apollon 10 pw laser: experimental and theoretical
  investigation of the temporal characteristics},\ }\href@noop {} {\bibfield
  {journal} {\bibinfo  {journal} {High Power Laser Science and Engineering}\
  }\textbf {\bibinfo {volume} {4}} (\bibinfo {year} {2016})}\BibitemShut
  {NoStop}%
\bibitem [{\citenamefont {Gauthier}\ \emph {et~al.}(2016)\citenamefont
  {Gauthier}, \citenamefont {Allaria}, \citenamefont {Coreno}, \citenamefont
  {Cudin}, \citenamefont {Dacasa}, \citenamefont {Danailov}, \citenamefont
  {Demidovich}, \citenamefont {Di~Mitri}, \citenamefont {Diviacco},
  \citenamefont {Ferrari} \emph {et~al.}}]{gauthier2016chirped}%
  \BibitemOpen
  \bibfield  {author} {\bibinfo {author} {\bibfnamefont {D.}~\bibnamefont
  {Gauthier}}, \bibinfo {author} {\bibfnamefont {E.}~\bibnamefont {Allaria}},
  \bibinfo {author} {\bibfnamefont {M.}~\bibnamefont {Coreno}}, \bibinfo
  {author} {\bibfnamefont {I.}~\bibnamefont {Cudin}}, \bibinfo {author}
  {\bibfnamefont {H.}~\bibnamefont {Dacasa}}, \bibinfo {author} {\bibfnamefont
  {M.~B.}\ \bibnamefont {Danailov}}, \bibinfo {author} {\bibfnamefont
  {A.}~\bibnamefont {Demidovich}}, \bibinfo {author} {\bibfnamefont
  {S.}~\bibnamefont {Di~Mitri}}, \bibinfo {author} {\bibfnamefont
  {B.}~\bibnamefont {Diviacco}}, \bibinfo {author} {\bibfnamefont
  {E.}~\bibnamefont {Ferrari}}, \emph {et~al.},\ }\bibfield  {title} {\bibinfo
  {title} {Chirped pulse amplification in an extreme-ultraviolet free-electron
  laser},\ }\href@noop {} {\bibfield  {journal} {\bibinfo  {journal} {Nature
  communications}\ }\textbf {\bibinfo {volume} {7}},\ \bibinfo {pages} {1}
  (\bibinfo {year} {2016})}\BibitemShut {NoStop}%
\bibitem [{\citenamefont {Pellegrini}(2000)}]{pellegrini2000high}%
  \BibitemOpen
  \bibfield  {author} {\bibinfo {author} {\bibfnamefont {C.}~\bibnamefont
  {Pellegrini}},\ }\bibfield  {title} {\bibinfo {title} {High power femtosecond
  pulses from an x-ray sase-fel},\ }\href@noop {} {\bibfield  {journal}
  {\bibinfo  {journal} {Nuclear Instruments and Methods in Physics Research
  Section A: Accelerators, Spectrometers, Detectors and Associated Equipment}\
  }\textbf {\bibinfo {volume} {445}},\ \bibinfo {pages} {124} (\bibinfo {year}
  {2000})}\BibitemShut {NoStop}%
\bibitem [{\citenamefont {Krinsky}\ and\ \citenamefont
  {Huang}(2003)}]{krinsky2003frequency}%
  \BibitemOpen
  \bibfield  {author} {\bibinfo {author} {\bibfnamefont {S.}~\bibnamefont
  {Krinsky}}\ and\ \bibinfo {author} {\bibfnamefont {Z.}~\bibnamefont
  {Huang}},\ }\bibfield  {title} {\bibinfo {title} {Frequency chirped
  self-amplified spontaneous-emission free-electron lasers},\ }\href@noop {}
  {\bibfield  {journal} {\bibinfo  {journal} {Physical Review Special
  Topics-Accelerators and Beams}\ }\textbf {\bibinfo {volume} {6}},\ \bibinfo
  {pages} {050702} (\bibinfo {year} {2003})}\BibitemShut {NoStop}%
\bibitem [{\citenamefont {Decker}\ \emph {et~al.}(2022)\citenamefont {Decker},
  \citenamefont {Bane}, \citenamefont {Colocho}, \citenamefont {Gilevich},
  \citenamefont {Marinelli}, \citenamefont {Sheppard}, \citenamefont {Turner},
  \citenamefont {Turner}, \citenamefont {Vetter}, \citenamefont {Halavanau}
  \emph {et~al.}}]{decker2022tunable}%
  \BibitemOpen
  \bibfield  {author} {\bibinfo {author} {\bibfnamefont {F.-J.}\ \bibnamefont
  {Decker}}, \bibinfo {author} {\bibfnamefont {K.~L.}\ \bibnamefont {Bane}},
  \bibinfo {author} {\bibfnamefont {W.}~\bibnamefont {Colocho}}, \bibinfo
  {author} {\bibfnamefont {S.}~\bibnamefont {Gilevich}}, \bibinfo {author}
  {\bibfnamefont {A.}~\bibnamefont {Marinelli}}, \bibinfo {author}
  {\bibfnamefont {J.~C.}\ \bibnamefont {Sheppard}}, \bibinfo {author}
  {\bibfnamefont {J.~L.}\ \bibnamefont {Turner}}, \bibinfo {author}
  {\bibfnamefont {J.~J.}\ \bibnamefont {Turner}}, \bibinfo {author}
  {\bibfnamefont {S.~L.}\ \bibnamefont {Vetter}}, \bibinfo {author}
  {\bibfnamefont {A.}~\bibnamefont {Halavanau}}, \emph {et~al.},\ }\bibfield
  {title} {\bibinfo {title} {Tunable x-ray free electron laser multi-pulses
  with nanosecond separation},\ }\href@noop {} {\bibfield  {journal} {\bibinfo
  {journal} {Scientific Reports}\ }\textbf {\bibinfo {volume} {12}},\ \bibinfo
  {pages} {1} (\bibinfo {year} {2022})}\BibitemShut {NoStop}%
\bibitem [{\citenamefont {Reiche}(1999)}]{reiche1999genesis}%
  \BibitemOpen
  \bibfield  {author} {\bibinfo {author} {\bibfnamefont {S.}~\bibnamefont
  {Reiche}},\ }\bibfield  {title} {\bibinfo {title} {Genesis 1.3: a fully 3d
  time-dependent fel simulation code},\ }\href@noop {} {\bibfield  {journal}
  {\bibinfo  {journal} {Nuclear Instruments and Methods in Physics Research
  Section A: Accelerators, Spectrometers, Detectors and Associated Equipment}\
  }\textbf {\bibinfo {volume} {429}},\ \bibinfo {pages} {243} (\bibinfo {year}
  {1999})}\BibitemShut {NoStop}%
\bibitem [{\citenamefont {Li}(2022)}]{li_2022}%
  \BibitemOpen
  \bibfield  {author} {\bibinfo {author} {\bibfnamefont {H.}~\bibnamefont
  {Li}},\ }\href {https://doi.org/10.6084/m9.figshare.19335638.v1} {\bibinfo
  {title} {Source code of xcpa}} (\bibinfo {year} {2022})\BibitemShut {NoStop}%
\bibitem [{\citenamefont {Batterman}\ and\ \citenamefont
  {Cole}(1964)}]{batterman1964dynamical}%
  \BibitemOpen
  \bibfield  {author} {\bibinfo {author} {\bibfnamefont {B.~W.}\ \bibnamefont
  {Batterman}}\ and\ \bibinfo {author} {\bibfnamefont {H.}~\bibnamefont
  {Cole}},\ }\bibfield  {title} {\bibinfo {title} {Dynamical diffraction of x
  rays by perfect crystals},\ }\href@noop {} {\bibfield  {journal} {\bibinfo
  {journal} {Reviews of modern physics}\ }\textbf {\bibinfo {volume} {36}},\
  \bibinfo {pages} {681} (\bibinfo {year} {1964})}\BibitemShut {NoStop}%
\bibitem [{\citenamefont {Martinez}\ \emph {et~al.}(1984)\citenamefont
  {Martinez}, \citenamefont {Gordon},\ and\ \citenamefont
  {Fork}}]{martinez1984negative}%
  \BibitemOpen
  \bibfield  {author} {\bibinfo {author} {\bibfnamefont {O.}~\bibnamefont
  {Martinez}}, \bibinfo {author} {\bibfnamefont {J.}~\bibnamefont {Gordon}},\
  and\ \bibinfo {author} {\bibfnamefont {R.}~\bibnamefont {Fork}},\ }\bibfield
  {title} {\bibinfo {title} {Negative group-velocity dispersion using
  refraction},\ }\href@noop {} {\bibfield  {journal} {\bibinfo  {journal} {JOSA
  A}\ }\textbf {\bibinfo {volume} {1}},\ \bibinfo {pages} {1003} (\bibinfo
  {year} {1984})}\BibitemShut {NoStop}%
\bibitem [{\citenamefont {Decker}\ \emph {et~al.}(2015)\citenamefont {Decker},
  \citenamefont {Gilevich}, \citenamefont {Huang}, \citenamefont {Loos},
  \citenamefont {Marinelli}, \citenamefont {Stan}, \citenamefont {Turner},
  \citenamefont {Van~Hoover}, \citenamefont {Vetter} \emph
  {et~al.}}]{decker2015two}%
  \BibitemOpen
  \bibfield  {author} {\bibinfo {author} {\bibfnamefont {F.}~\bibnamefont
  {Decker}}, \bibinfo {author} {\bibfnamefont {S.}~\bibnamefont {Gilevich}},
  \bibinfo {author} {\bibfnamefont {Z.}~\bibnamefont {Huang}}, \bibinfo
  {author} {\bibfnamefont {H.}~\bibnamefont {Loos}}, \bibinfo {author}
  {\bibfnamefont {A.}~\bibnamefont {Marinelli}}, \bibinfo {author}
  {\bibfnamefont {C.}~\bibnamefont {Stan}}, \bibinfo {author} {\bibfnamefont
  {J.}~\bibnamefont {Turner}}, \bibinfo {author} {\bibfnamefont
  {Z.}~\bibnamefont {Van~Hoover}}, \bibinfo {author} {\bibfnamefont
  {S.}~\bibnamefont {Vetter}}, \emph {et~al.},\ }\bibfield  {title} {\bibinfo
  {title} {Two bunches with ns-separation with lcls},\ }in\ \href@noop {}
  {\emph {\bibinfo {booktitle} {Proceedings of the 37th International Free
  Electron Laser Conference (FEL 2015), HS Kang, D.-E. Kim, VRW Schaa,
  Eds.(JACoW, Geneva, Switzerland, 2015), p. WEP023}}}\ (\bibinfo {year}
  {2015})\BibitemShut {NoStop}%
\bibitem [{\citenamefont {Martinez}(1987)}]{martinez19873000}%
  \BibitemOpen
  \bibfield  {author} {\bibinfo {author} {\bibfnamefont {O.}~\bibnamefont
  {Martinez}},\ }\bibfield  {title} {\bibinfo {title} {3000 times grating
  compressor with positive group velocity dispersion: Application to fiber
  compensation in 1.3-1.6 $\mu$m region},\ }\href@noop {} {\bibfield  {journal}
  {\bibinfo  {journal} {IEEE Journal of Quantum Electronics}\ }\textbf
  {\bibinfo {volume} {23}},\ \bibinfo {pages} {59} (\bibinfo {year}
  {1987})}\BibitemShut {NoStop}%
\bibitem [{\citenamefont {Schwartz}\ \emph {et~al.}(2021)\citenamefont
  {Schwartz}, \citenamefont {Raj}, \citenamefont {Jamnuch}, \citenamefont
  {Hull}, \citenamefont {Miotti}, \citenamefont {Lam}, \citenamefont
  {Nordlund}, \citenamefont {Uzundal}, \citenamefont {Das~Pemmaraju},
  \citenamefont {Mincigrucci}, \citenamefont {Foglia}, \citenamefont
  {Simoncig}, \citenamefont {Coreno}, \citenamefont {Masciovecchio},
  \citenamefont {Giannessi}, \citenamefont {Poletto}, \citenamefont {Principi},
  \citenamefont {Zuerch}, \citenamefont {Pascal}, \citenamefont {Drisdell},\
  and\ \citenamefont {Saykally}}]{PhysRevLett.127.096801}%
  \BibitemOpen
  \bibfield  {author} {\bibinfo {author} {\bibfnamefont {C.~P.}\ \bibnamefont
  {Schwartz}}, \bibinfo {author} {\bibfnamefont {S.~L.}\ \bibnamefont {Raj}},
  \bibinfo {author} {\bibfnamefont {S.}~\bibnamefont {Jamnuch}}, \bibinfo
  {author} {\bibfnamefont {C.~J.}\ \bibnamefont {Hull}}, \bibinfo {author}
  {\bibfnamefont {P.}~\bibnamefont {Miotti}}, \bibinfo {author} {\bibfnamefont
  {R.~K.}\ \bibnamefont {Lam}}, \bibinfo {author} {\bibfnamefont
  {D.}~\bibnamefont {Nordlund}}, \bibinfo {author} {\bibfnamefont {C.~B.}\
  \bibnamefont {Uzundal}}, \bibinfo {author} {\bibfnamefont {C.}~\bibnamefont
  {Das~Pemmaraju}}, \bibinfo {author} {\bibfnamefont {R.}~\bibnamefont
  {Mincigrucci}}, \bibinfo {author} {\bibfnamefont {L.}~\bibnamefont {Foglia}},
  \bibinfo {author} {\bibfnamefont {A.}~\bibnamefont {Simoncig}}, \bibinfo
  {author} {\bibfnamefont {M.}~\bibnamefont {Coreno}}, \bibinfo {author}
  {\bibfnamefont {C.}~\bibnamefont {Masciovecchio}}, \bibinfo {author}
  {\bibfnamefont {L.}~\bibnamefont {Giannessi}}, \bibinfo {author}
  {\bibfnamefont {L.}~\bibnamefont {Poletto}}, \bibinfo {author} {\bibfnamefont
  {E.}~\bibnamefont {Principi}}, \bibinfo {author} {\bibfnamefont
  {M.}~\bibnamefont {Zuerch}}, \bibinfo {author} {\bibfnamefont {T.~A.}\
  \bibnamefont {Pascal}}, \bibinfo {author} {\bibfnamefont {W.~S.}\
  \bibnamefont {Drisdell}},\ and\ \bibinfo {author} {\bibfnamefont {R.~J.}\
  \bibnamefont {Saykally}},\ }\bibfield  {title} {\bibinfo {title}
  {Angstrom-resolved interfacial structure in buried organic-inorganic
  junctions},\ }\href {https://doi.org/10.1103/PhysRevLett.127.096801}
  {\bibfield  {journal} {\bibinfo  {journal} {Phys. Rev. Lett.}\ }\textbf
  {\bibinfo {volume} {127}},\ \bibinfo {pages} {096801} (\bibinfo {year}
  {2021})}\BibitemShut {NoStop}%
\bibitem [{\citenamefont {Berger}\ \emph {et~al.}(2021)\citenamefont {Berger},
  \citenamefont {Jamnuch}, \citenamefont {Uzundal}, \citenamefont {Woodahl},
  \citenamefont {Padmanabhan}, \citenamefont {Amado}, \citenamefont {Manset},
  \citenamefont {Hirata}, \citenamefont {Kubota}, \citenamefont {Owada} \emph
  {et~al.}}]{berger2021extreme}%
  \BibitemOpen
  \bibfield  {author} {\bibinfo {author} {\bibfnamefont {E.}~\bibnamefont
  {Berger}}, \bibinfo {author} {\bibfnamefont {S.}~\bibnamefont {Jamnuch}},
  \bibinfo {author} {\bibfnamefont {C.~B.}\ \bibnamefont {Uzundal}}, \bibinfo
  {author} {\bibfnamefont {C.}~\bibnamefont {Woodahl}}, \bibinfo {author}
  {\bibfnamefont {H.}~\bibnamefont {Padmanabhan}}, \bibinfo {author}
  {\bibfnamefont {A.}~\bibnamefont {Amado}}, \bibinfo {author} {\bibfnamefont
  {P.}~\bibnamefont {Manset}}, \bibinfo {author} {\bibfnamefont
  {Y.}~\bibnamefont {Hirata}}, \bibinfo {author} {\bibfnamefont
  {Y.}~\bibnamefont {Kubota}}, \bibinfo {author} {\bibfnamefont
  {S.}~\bibnamefont {Owada}}, \emph {et~al.},\ }\bibfield  {title} {\bibinfo
  {title} {Extreme ultraviolet second harmonic generation spectroscopy in a
  polar metal},\ }\href@noop {} {\bibfield  {journal} {\bibinfo  {journal}
  {Nano letters}\ }\textbf {\bibinfo {volume} {21}},\ \bibinfo {pages} {6095}
  (\bibinfo {year} {2021})}\BibitemShut {NoStop}%
\bibitem [{\citenamefont {Uzundal}\ \emph {et~al.}(2021)\citenamefont
  {Uzundal}, \citenamefont {Jamnuch}, \citenamefont {Berger}, \citenamefont
  {Woodahl}, \citenamefont {Manset}, \citenamefont {Hirata}, \citenamefont
  {Sumi}, \citenamefont {Amado}, \citenamefont {Akai}, \citenamefont {Kubota},
  \citenamefont {Owada}, \citenamefont {Tono}, \citenamefont {Yabashi},
  \citenamefont {Freeland}, \citenamefont {Schwartz}, \citenamefont {Drisdell},
  \citenamefont {Matsuda}, \citenamefont {Pascal}, \citenamefont {Zong},\ and\
  \citenamefont {Zuerch}}]{PhysRevLett.127.237402}%
  \BibitemOpen
  \bibfield  {author} {\bibinfo {author} {\bibfnamefont {C.~B.}\ \bibnamefont
  {Uzundal}}, \bibinfo {author} {\bibfnamefont {S.}~\bibnamefont {Jamnuch}},
  \bibinfo {author} {\bibfnamefont {E.}~\bibnamefont {Berger}}, \bibinfo
  {author} {\bibfnamefont {C.}~\bibnamefont {Woodahl}}, \bibinfo {author}
  {\bibfnamefont {P.}~\bibnamefont {Manset}}, \bibinfo {author} {\bibfnamefont
  {Y.}~\bibnamefont {Hirata}}, \bibinfo {author} {\bibfnamefont
  {T.}~\bibnamefont {Sumi}}, \bibinfo {author} {\bibfnamefont {A.}~\bibnamefont
  {Amado}}, \bibinfo {author} {\bibfnamefont {H.}~\bibnamefont {Akai}},
  \bibinfo {author} {\bibfnamefont {Y.}~\bibnamefont {Kubota}}, \bibinfo
  {author} {\bibfnamefont {S.}~\bibnamefont {Owada}}, \bibinfo {author}
  {\bibfnamefont {K.}~\bibnamefont {Tono}}, \bibinfo {author} {\bibfnamefont
  {M.}~\bibnamefont {Yabashi}}, \bibinfo {author} {\bibfnamefont {J.~W.}\
  \bibnamefont {Freeland}}, \bibinfo {author} {\bibfnamefont {C.~P.}\
  \bibnamefont {Schwartz}}, \bibinfo {author} {\bibfnamefont {W.~S.}\
  \bibnamefont {Drisdell}}, \bibinfo {author} {\bibfnamefont {I.}~\bibnamefont
  {Matsuda}}, \bibinfo {author} {\bibfnamefont {T.~A.}\ \bibnamefont {Pascal}},
  \bibinfo {author} {\bibfnamefont {A.}~\bibnamefont {Zong}},\ and\ \bibinfo
  {author} {\bibfnamefont {M.}~\bibnamefont {Zuerch}},\ }\bibfield  {title}
  {\bibinfo {title} {Polarization-resolved extreme-ultraviolet second-harmonic
  generation from ${\mathrm{linbo}}_{3}$},\ }\href
  {https://doi.org/10.1103/PhysRevLett.127.237402} {\bibfield  {journal}
  {\bibinfo  {journal} {Phys. Rev. Lett.}\ }\textbf {\bibinfo {volume} {127}},\
  \bibinfo {pages} {237402} (\bibinfo {year} {2021})}\BibitemShut {NoStop}%
\bibitem [{\citenamefont {Tanaka}(2003)}]{tanaka2003ultrafast}%
  \BibitemOpen
  \bibfield  {author} {\bibinfo {author} {\bibfnamefont {S.}~\bibnamefont
  {Tanaka}},\ }\bibfield  {title} {\bibinfo {title} {Ultrafast relaxation
  dynamics of the one-dimensional molecular chain: The time-resolved
  spontaneous emission and exciton coherence},\ }\href@noop {} {\bibfield
  {journal} {\bibinfo  {journal} {The Journal of chemical physics}\ }\textbf
  {\bibinfo {volume} {119}},\ \bibinfo {pages} {4891} (\bibinfo {year}
  {2003})}\BibitemShut {NoStop}%
\bibitem [{\citenamefont {Tanaka}\ and\ \citenamefont
  {Mukamel}(2004)}]{tanaka2004simulation}%
  \BibitemOpen
  \bibfield  {author} {\bibinfo {author} {\bibfnamefont {S.}~\bibnamefont
  {Tanaka}}\ and\ \bibinfo {author} {\bibfnamefont {S.}~\bibnamefont
  {Mukamel}},\ }\bibfield  {title} {\bibinfo {title} {Simulation of optical and
  x-ray sum frequency generation spectroscopy of one-dimensional molecular
  chain},\ }\href@noop {} {\bibfield  {journal} {\bibinfo  {journal} {Journal
  of electron spectroscopy and related phenomena}\ }\textbf {\bibinfo {volume}
  {136}},\ \bibinfo {pages} {185} (\bibinfo {year} {2004})}\BibitemShut
  {NoStop}%
\bibitem [{\citenamefont {Pandey}\ and\ \citenamefont
  {Mukamel}(2006)}]{pandey2006simulation}%
  \BibitemOpen
  \bibfield  {author} {\bibinfo {author} {\bibfnamefont {R.}~\bibnamefont
  {Pandey}}\ and\ \bibinfo {author} {\bibfnamefont {S.}~\bibnamefont
  {Mukamel}},\ }\bibfield  {title} {\bibinfo {title} {Simulation of x-ray
  absorption near-edge spectra and x-ray fluorescence spectra of optically
  excited molecules},\ }\href@noop {} {\bibfield  {journal} {\bibinfo
  {journal} {The Journal of chemical physics}\ }\textbf {\bibinfo {volume}
  {124}},\ \bibinfo {pages} {094106} (\bibinfo {year} {2006})}\BibitemShut
  {NoStop}%
\bibitem [{\citenamefont {Campbell}\ and\ \citenamefont
  {Mukamel}(2004)}]{campbell2004simulation}%
  \BibitemOpen
  \bibfield  {author} {\bibinfo {author} {\bibfnamefont {L.}~\bibnamefont
  {Campbell}}\ and\ \bibinfo {author} {\bibfnamefont {S.}~\bibnamefont
  {Mukamel}},\ }\bibfield  {title} {\bibinfo {title} {Simulation of x-ray
  absorption near edge spectra of electronically excited ruthenium tris-2,
  2$'$-bipyridine},\ }\href@noop {} {\bibfield  {journal} {\bibinfo  {journal}
  {The Journal of chemical physics}\ }\textbf {\bibinfo {volume} {121}},\
  \bibinfo {pages} {12323} (\bibinfo {year} {2004})}\BibitemShut {NoStop}%
\bibitem [{\citenamefont {Berman}\ and\ \citenamefont
  {Mukamel}(2004)}]{PhysRevB.69.155430}%
  \BibitemOpen
  \bibfield  {author} {\bibinfo {author} {\bibfnamefont {O.}~\bibnamefont
  {Berman}}\ and\ \bibinfo {author} {\bibfnamefont {S.}~\bibnamefont
  {Mukamel}},\ }\bibfield  {title} {\bibinfo {title} {Current profiles of
  molecular nanowires: Density-functional theory green's function
  representation},\ }\href {https://doi.org/10.1103/PhysRevB.69.155430}
  {\bibfield  {journal} {\bibinfo  {journal} {Phys. Rev. B}\ }\textbf {\bibinfo
  {volume} {69}},\ \bibinfo {pages} {155430} (\bibinfo {year}
  {2004})}\BibitemShut {NoStop}%
\bibitem [{\citenamefont {Shvyd'Ko}(2004)}]{shvyd2004x}%
  \BibitemOpen
  \bibfield  {author} {\bibinfo {author} {\bibfnamefont {Y.}~\bibnamefont
  {Shvyd'Ko}},\ }\href@noop {} {\emph {\bibinfo {title} {X-ray optics:
  high-energy-resolution applications}}},\ Vol.~\bibinfo {volume} {98}\
  (\bibinfo  {publisher} {Springer Science \& Business Media},\ \bibinfo {year}
  {2004})\BibitemShut {NoStop}%
\end{thebibliography}
%

\appendix

\section{Ray-tracing analysis of the Stretcher} \label{appendix:RayTracingStretcher}
The ray-tracing analysis of the stretcher is presented in this appendix.
Symbols in bold font represent vectors.

\begin{figure}[bht!]
    \centering
    \includegraphics[width=0.3\textwidth]{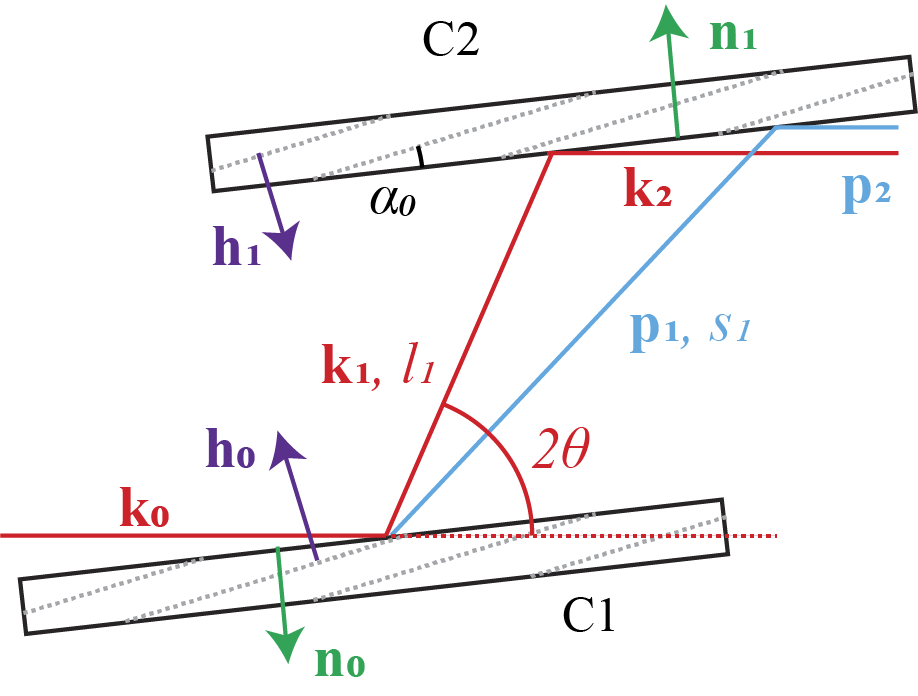}
    \caption{Schematics of the left half of the stretcher.}
    \label{fig:stretcherRayTracing}
\end{figure}

Because the stretcher is mirror symmetric with respect to its central plane, one only needs to analyze half of it.
The left half of the stretcher and variables required for the derivation is visualized in Figure~\ref{fig:stretcherRayTracing}.

In Figure~\ref{fig:stretcherRayTracing}, $\textbf{k}_0$, $\textbf{k}_1$, $\textbf{k}_2$ stand for the unit vector of the propagation direction of the reference photon. 
The $\textbf{p}_1$ and $\textbf{p}_2$ stand for the unit vector of the propagation direction of a photon with a different photon energy and the same incident direction. 
The Figure~\ref{fig:stretcherRayTracing} shows the case where $\textbf{p}_1$ corresponds to higher photon energy with respect to $\textbf{k}_1$.
The symbol $l_1$ and $s_1$ stand for the length of the line section joining the two crystals $C1$ and $C2$.

The path-length difference of $\textbf{p}_1$ with respect to $\textbf{k}_1$ is
\begin{equation}
    \delta\left(L\right) =2s_1-2l_1 -2\left(s_1p_1-l_1k_1\right)\cdot k_0 .
    \label{eq:stretcherPL0}
\end{equation}
Because both $l_1 \textbf{k}_1$ and $s_1 \textbf{p}_1$ are on the same reflection surface of crystal $C2$, one can represent 
$s_1$ with other quantities:
\begin{equation}
    s_1=\frac{\textbf{k}_1\cdot \textbf{n}_1}{\textbf{p}_1\cdot \textbf{n}_1}\ l_1 .
\end{equation}
Therefore, one can simplify equation (\ref{eq:stretcherPL0}) to 
\begin{equation}
    \delta\left(L\right) =2\left(\left(1-\textbf{p}_1\cdot \textbf{p}_0\right)\frac{\textbf{k}_1\cdot \textbf{n}_1}{\textbf{p}_1\cdot \textbf{n}_1}+\textbf{k}_1\cdot \textbf{k}_0-1\right)l_1 .
    \label{eq:stretcherPL1}
\end{equation}

The energy dependence enters equation (\ref{eq:stretcherPL1}) through $\textbf{p}_1$.
Define wave-vectors corresponding to $\textbf{k}_0$, $\textbf{k}_1$, $\textbf{p}_0$ and $\textbf{p}_1$ as $\textbf{K}_0$, $\textbf{K}_1$, $\textbf{P}_0$ and $\textbf{P}_1$, then the following relation holds:
\begin{gather}
    \textbf{k}_i = \textbf{K}_i |\textbf{K}_i|^{-1},\ \ \ i=0,1 , \\
    \textbf{p}_i = \textbf{P}_i |\textbf{P}_i|^{-1},\ \ \ i=0,1  , \label{eq:p1} \\
    \textbf{p}_i = \textbf{k}_i + \delta(\textbf{k}_i),\ \ \ i=0,1  ,\\
    \textbf{P}_i = \textbf{K}_i + \delta(\textbf{K}_i),\ \ \ i=0,1 .
\end{gather}
According to the dynamical diffraction theory \cite{shvyd2004x},
\begin{gather}
    \textbf{K}_1=\textbf{K}_0+\textbf{h}_0+\Delta_0\textbf{n}_0  , \\
    |\textbf{K}_1|^2 = |\textbf{K}_0|^2  .
    \label{eq:DDT}
\end{gather}
where $\textbf{h}_0$ is the reciprocal lattice of the $C1$ Bragg reflection. 

If one change $\textbf{K}_0$ to $\textbf{P}_0=\textbf{K}_0 + \delta(\textbf{K}_0)$, then $\Delta_0$ changes to $\Delta_0+ \delta(\Delta_0)$.
One can derive from equation (\ref{eq:DDT}) that:
\begin{equation}
    \delta\left(\Delta_0\right)=\frac{\left(\textbf{K}_0-\textbf{K}_1\right)\cdot\delta\left(\textbf{K}_0\right)}{\textbf{K}_1\cdot \textbf{n}_0} .
\end{equation}
Therefore
\begin{equation}
    \delta\left(\textbf{K}_1\right)=\delta\left(\textbf{K}_0\right)+\frac{\left(\textbf{K}_0-\textbf{K}_1\right)\cdot\delta\left(\textbf{K}_0\right)}{\textbf{K}_1\cdot \textbf{n}_0}\textbf{n}_0  .
\end{equation}
In the following, we are only interested in the energy change. 
Therefore, assume that 
\begin{equation}
    \delta\left(\textbf{K}_0\right)=\frac{\textbf{K}_0}{|\textbf{K}_0|}\frac{\delta E}{\hbar} = \textbf{k}_0 \frac{\delta E}{\hbar} ,
\end{equation}
where $\hbar$ is the reduced Planck constant.
Therefore
\begin{equation}
    \delta\left(\textbf{K}_1\right)=\frac{\delta E}{\hbar}\left(\textbf{k}_0+\frac{\left(\textbf{k}_0-\textbf{k}_1\right)\cdot \textbf{k}_0}{\textbf{k}_1\cdot \textbf{n}_0}\textbf{n}_0\right) .
    \label{eq:deltaK1}
\end{equation}
The dependence of $\textbf{p}_1$ on the energy change $\delta E$ can be obtained from equation (\ref{eq:p1}) and (\ref{eq:deltaK1})
\begin{equation}
    \textbf{p}_1\approx \textbf{k}_1+\frac{\delta E}{E_0}\left(\textbf{k}_0-\textbf{k}_1+\frac{\left(\textbf{k}_0-\textbf{k}_1\right)\cdot \textbf{k}_0}{\textbf{k}_1\cdot \textbf{n}_0}\textbf{n}_0\right) .
    \label{eq:p1vsEnergy}
\end{equation}

\begin{figure*}[bht!]
    \centering
    \includegraphics[width=0.8\textwidth]{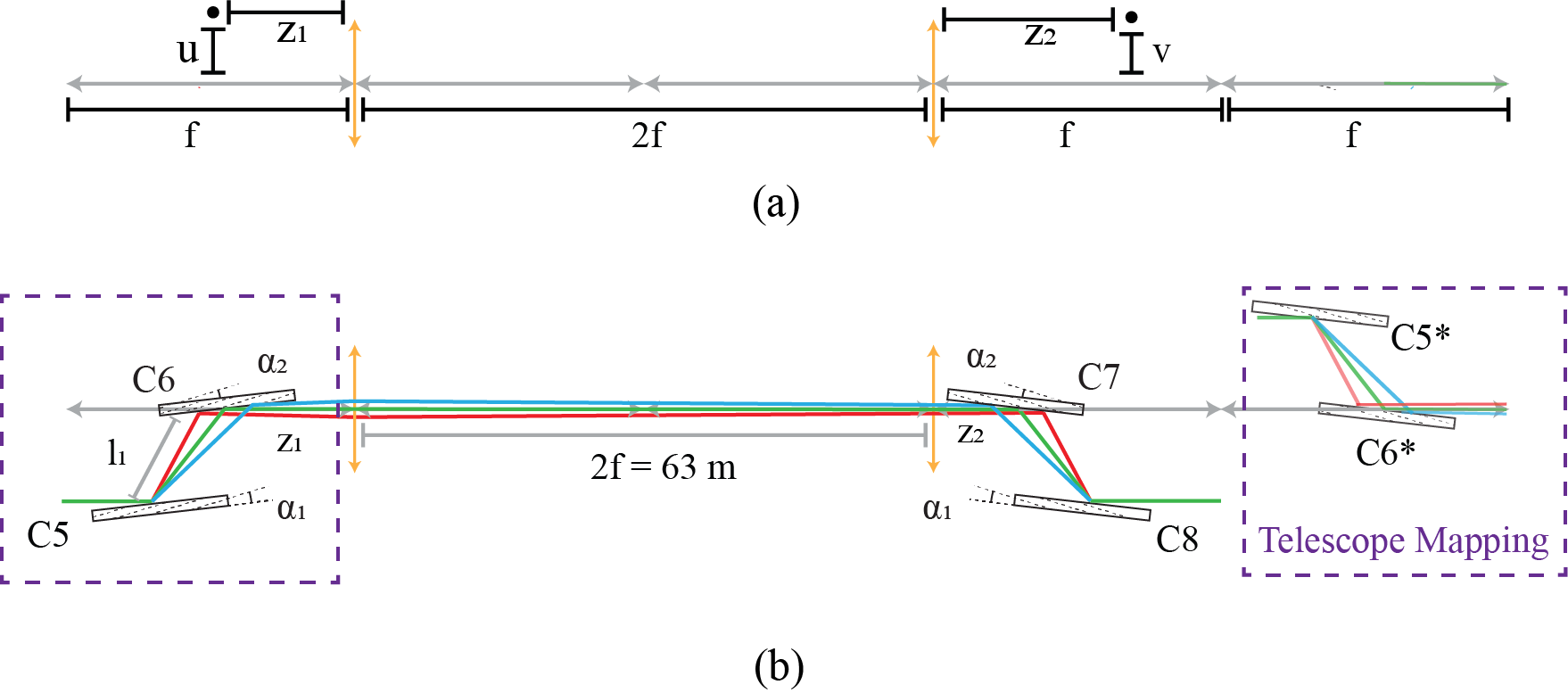}
    \caption{
    (a) Definition of variables for the 2f imaging system.
    (b) The $C5^*$ and $C6^*$ are the image of the crystal pair $C5$-$C6$ produced by the telescope with mapping defined through equation (\ref{eq:telescope mapping 1}) and (\ref{eq:telescope mapping 2}).
    They are constructed by keeping the direction along the telescope's optical axis unchanged and applying central inversion along directions perpendicular to the optical axis.
    The $z_1$ and $z_2$ are the distances of $C6$ and $C7$ with their adjacent focusing lens respectively. 
    In the body text, we have assumed that they are negligible.
    }
    \label{fig:imaging}
\end{figure*}

With equation (\ref{eq:stretcherPL1}) and (\ref{eq:p1vsEnergy}), one obtains an explicit formula of the dependence of the path length on the energy $\delta E$. 
We use \emph{Mathematica} to simplify the derivation. 

The Bragg angle $\theta$ is defined through
\begin{equation}
    \textbf{k}_0\cdot \textbf{k}_1 = \cos{2\theta}.
    \label{eq:theta}
\end{equation}
When the asymmetry angle is $\alpha_0$, $\textbf{k}_0$, $\textbf{k}_1$ and $\textbf{n}_0$ have the following expression:
\begin{align}
    \textbf{k}_0 &= (1,~0) ,\\
    \textbf{k}_1 &= \left(\cos{2\theta},~ \sin{2\theta}\right) ,\\
    \textbf{n}_0 &= \left(-\sin{(\theta-\alpha_0)}, ~ \cos{(\theta-\alpha_0)}\right) .
\end{align}
With \emph{Mathematica}, one can easily obtain the following simplified expression for equation (\ref{eq:stretcherPL1}):
\begin{equation}
    \delta\left(L\right)=-8\frac{\sin^2{\alpha_0}\sin^2{\theta}}{\sin^2{\left(\alpha_0+\theta\right)}}l_1\times\frac{\delta E}{E_0}
    \label{eq:stretcherPL2}
\end{equation}

According to the geometry, when the gap size of the channel-cut is $d$, 
\begin{equation}
    l_1 = \frac{d}{\sin{(\theta + \alpha_0)}}
    \label{eq:l1 and d}
\end{equation}
With equation (\ref{eq:stretcherPL2}) and (\ref{eq:l1 and d}), one obtains the equation (\ref{eq:stretcherPL}).

\section{Ray-tracing analysis of the Compressor}\label{appendix:RayTracingCompressor}

We divide the light path difference of the compressor between $C5$ and $C7$ into 3 parts. We use $C5$-$C6$ for the path length between $C5$ and $C6$ shown in Figure~\ref{fig:imaging} (b), $C6$-$C7$ for the light path length between $C6$ and $C7$ and $C7$-$C8$ for the light path length between $C7$ and $C8$.

The $C5$-$C6$ and $C7$-$C8$ follow the same analysis method as that of the stretcher.
Because there is a telescope laying in between crystal $C6$ and $C7$, the $C6$-$C7$ is more complicated to analyze. 
To calculate $C6$-$C7$, we borrow the idea of the first analysis of an optical system of this kind \cite{martinez19873000}. 
We summarize the details in the next subsection, \ref{subsection:imaging}.

To facilitate the verification of the derivation, we divide our analysis into 3 parts.
In subsection \ref{subsection:imaging}, we show how we construct the image $C5^*$ and $C6^*$ with Newton's equation of thin lens.
Then we present how we combine this imaging process with the analysis technique presented in Ref \cite{martinez19873000}.
In subsection \ref{subsection:Definition of variables}, we give the explicit definition of the variables we used to derive the light path length difference of the compressor.
In subsection \ref{subsection:calculation}, we present the derivation of the equation \ref{eq:compressorPL} with variables defined in subsection \ref{subsection:Definition of variables}.

\begin{figure*}[bht!]
    \centering
    \includegraphics[width=0.8\textwidth]{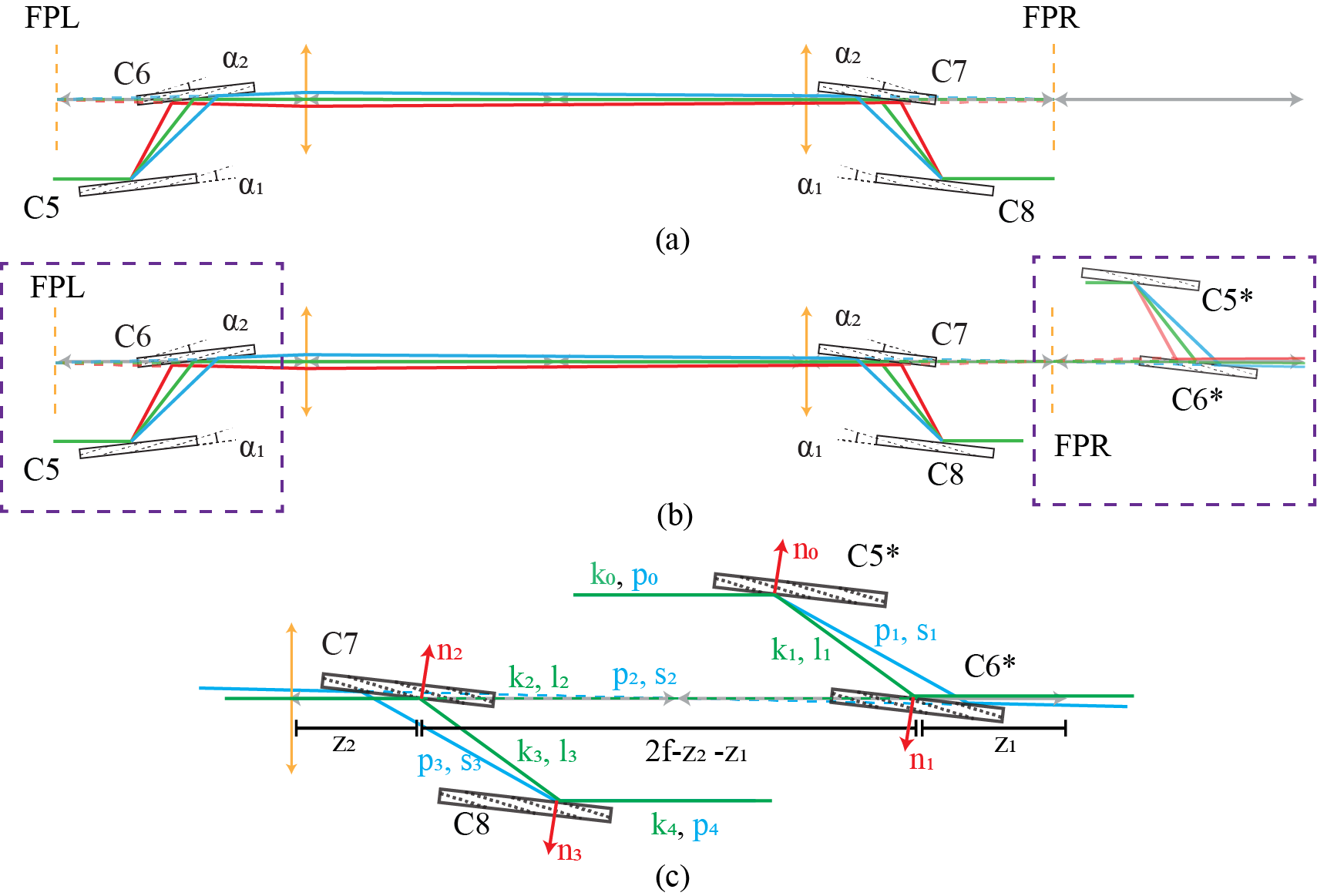}
    \caption{
    (a) Illustration of the negative path length technique described in Ref \cite{martinez19873000}. The dashed lines are for the backwards propagations from $C6$ and FPR to FPL and $C7$ respectively.
    (b) Illustration showing that the summation of $C6$-FPL and FPR-$C7$ is equivalent to the path length connecting $C6^*$ and $C7$ along the negative propagation direction of the image beam reflected from $C6^*$. The dashed lines stand for these negative propagation paths.
    (c) The $C5^*$-$C6^*$-$C7$-$C8$ region with all variables defined for the concrete calculation. The concrete meaning of these variables are specified in subsection \ref{subsection:Definition of variables}. The dashed lines stand for negative propagation paths. The green lines are for the reference photon energy and blue lines are for a different photon energy to be calculated.
    }
    \label{fig:compressor setup}
\end{figure*}

\subsection{Imaging}\label{subsection:imaging}

With Newton's equation of thin lens, one can easily verify that, for an object at position $(u, z_1)$, the image is at position $(v, z_2)$
\begin{gather}
   v = -u  \label{eq:telescope mapping 1}\\
   z_2 = 2f -z_1 \label{eq:telescope mapping 2}
\end{gather}
Definitions of these quantities are shown in Figure~\ref{fig:imaging} (a).

According to the mapping equation (\ref{eq:telescope mapping 1}) and (\ref{eq:telescope mapping 2}), one can construct the image of $C5$ and $C6$ on the right of the telescope as $C5^*$ and $C6^*$, by keeping the direction along the telescope's optical axis unchanged and applying central inversion along directions perpendicular to the optical axis. 
This imaging process is illustrated with Figure~\ref{fig:imaging} (b).

This mapping with equation (\ref{eq:telescope mapping 1}) and (\ref{eq:telescope mapping 2}) provides an extension of the technique described in Ref \cite{martinez19873000} to analyze the light path length difference of a telescope and two dispersive optics, which in our case are $C6$ and $C7$. 

The technique described in Ref \cite{martinez19873000}, goes as the following. First propagate the reflected ray from $C6$ along its negative direction to the focal plane on the left of the first lens (FPL) and calculate the negative value of this path length. Then apply an central inversion of the ray on FPL, which gives the light field on the focal plane to the right of the second focusing lens (FPR).
Then calculate the negative path length between FPR and $C7$.
These backwards propagation of $C6$-FPL and FPR-$C7$ are illustrated with dashed lines in Figure~\ref{fig:compressor setup} (a) with the FPL and FPR explicitly defined in the plot.
It is proved in Ref \cite{martinez19873000} that the path length difference for different photon energies between the two dispersive optics, $C6$ and $C7$, is the same as the corresponding difference of summation of the two negative path lengths, i.e. $C6$-FPL and FPR-$C7$.

To combine the negative path length technique with the imaging of the telescope, one can verify that the mapping described in the previous paragraph (also the one in Ref \cite{martinez19873000}) is a special case of the mapping with the telescope represented with equation (\ref{eq:telescope mapping 1}) and (\ref{eq:telescope mapping 2}), where $z_1 = f$. 
Therefore, the total negative path length of $C6$-FPL and FPR-$C7$ is the same as the negative path length connecting $C6^*$ and $C7$ along the negative propagation direction of the image beam reflected from $C6^*$, as is illustrated in Figure~\ref{fig:compressor setup} (b) with dashed lines.

According to the mapping equations (\ref{eq:telescope mapping 1}) and (\ref{eq:telescope mapping 2}), the light path $C5$-$C6$ is the same as $C5^*$-$C6^*$ in the image region.
Combining this with the analysis of $C6^*$-$C7$ presented above, we can conclude that to calculate the light path length difference for photons with different energies in this system, we only need to calculate the path length difference in among $C5^*$-$C6^*$-$C7$-$C8$, which is shown in details in Figure~\ref{fig:compressor setup} (c).

\subsection{Definition of variables}\label{subsection:Definition of variables}

Previously, we have shown that, one only needs to analyze the $C5^*$-$C6^*$-$C7$-$C8$ light path shown in Figure~\ref{fig:compressor setup} (c).
In this section, we present the definition of the quantities defined in Figure~\ref{fig:compressor setup} (c), which will be used to perform the concrete derivation in the next subsection. 

In Figure~\ref{fig:compressor setup} (c), $\textbf{k}_i$, $i=0,1,2,3,4$, are unit vectors for propagation directions of the reference photon, $l_i$, $i=1,2,3$, are light path lengths of the reference photon between different crystals, and $\textbf{p}_i$, $i=0,1,2,3,4$, and $s_i$, $i=1,2,3$, are the corresponding quantities for another photon with the same incident angle and a different photon energy.
The $\textbf{n}_i$, $i=0,1,2,3$, are the normal directions of reflection surfaces of different crystals.

Define the Bragg angle of the reference photon for $C5^*$ and $C6^*$ as 
\begin{align}
    \textbf{k}_0\cdot \textbf{k}_1 &= \cos{2\theta_1} , \label{eq:theta1} \\
    \textbf{k}_1\cdot \textbf{k}_2 &= \cos{2\theta_2} . \label{eq:theta2}
\end{align}
We define these Bragg angles because the asymmetry angles $\alpha_0\neq \alpha_1 \neq \alpha_2$, and the asymmetry angle changes the Bragg angle \cite{shvyd2004x} slightly and therefore, strictly speaking, $\theta \neq \theta_1 \neq \theta_2$.
In the main body of the derivation, we keep these strict definition to ensure an accurate estimation of the path length difference. 
We only use the approximation, $\theta \approx \theta_1 \approx \theta_2$, in the end to derive the equation (\ref{eq:compressorPL}) in the body text.

To simplify the calculation, the x-axis is aligned with $\textbf{k}_2$. 
In this case, the vectors $\textbf{k}_i$, $i=0,1,2$, in Figure~\ref{fig:compressor setup} (c) can be explicitly defined as:
\begin{align}
    \textbf{k}_0&=\left(\cos{\left(2\theta_1-2\theta_2\right)},\ \sin{\left(2\theta_1-2\theta_2\right)}\right) =\textbf{k}_4 , \\
    \textbf{k}_1&=\left(\cos{\left(2\theta_2\right)},\ -\sin{\left(2\theta_2\right)}\right) =\textbf{k}_3 ,\\
    \textbf{k}_2&=\left(-1,\ 0\right).
\end{align}
The normal directions $\textbf{n}_i$, $i=0,1,2,3$, have the expression:
\begin{align}
    \textbf{n}_0&=\left(\sin{\left(\theta_2-\alpha_1\right)},\cos{\left(\theta_2-\alpha_1\right)}\right) =-\textbf{n}_3,\\
    \textbf{n}_1&=\left(\sin{\left(\theta_2-\alpha_2\right)},\cos{\left(\theta_2-\alpha_2\right)}\right) = -\textbf{n}_2 .
\end{align}
In the numerical simulation, we set $l_1=l_3$ for the central wave-vector. 
Therefore, we make the same assumption here.
According to the geometry, $l_2 = 2f - z_1 - z_2$.

\subsection{Light path length calculation}\label{subsection:calculation}
In this section, we implement the calculation describe in subsection \ref{subsection:imaging}, with the variables define in the previous subsection.

The path length of the blue lines in Figure~\ref{fig:compressor setup} (c) with respect to the reference green line can be expressed as:
\begin{multline}
    \delta(L) =s_1-s_2+s_3 - \left(s_1\textbf{p}_1+s_2\textbf{p}_2+s_3p_3\right)\cdot \textbf{k}_4  \\
     - l_1 + l_2 - l_3 + \left(l_1\textbf{k}_1+l_2\textbf{k}_2+l_3\textbf{k}_3\right)\cdot \textbf{k}_4
     \label{eq:RayTracingCompressor1}
\end{multline}

In equation (\ref{eq:RayTracingCompressor1}), the $s_i$, $i=1,2,3$ can be represented with $\textbf{p}_i$, $\textbf{k}_i$ and $l_i$:
\begin{align}
    s_1&=\frac{\textbf{k}_1\cdot \textbf{n}_1}{\textbf{p}_1\cdot \textbf{n}_1}\ l_1 ,\\
    s_2&=\frac{\left(l_1\textbf{k}_1+l_2\textbf{k}_2\right)\cdot \textbf{n}_2-s_1\textbf{p}_1\cdot \textbf{n}_1}{\textbf{p}_2\cdot \textbf{n}_2} ,\\
    s_3&=\frac{\left(l_1\textbf{k}_1+l_2\textbf{k}_2+l_3\textbf{k}_3\right)\cdot \textbf{n}_3-\left(s_1\textbf{p}_1+s_2\textbf{p}_2\right)\cdot \textbf{n}_3}{p_3\cdot \textbf{n}_3} .
\end{align}

Then, the energy dependence enters equation (\ref{eq:RayTracingCompressor1}) through $\textbf{p}_i$, $i=1,2,3$.
The $C6^*$-$C7$ crystal pair does not change the wave-vector.
Therefore, $\textbf{p}_3 = \textbf{p}_1$. 
We need to find the energy dependence of $\textbf{p}_1$ and $\textbf{p}_2$.

The energy dependence of $\textbf{p}_1$ is the same as that in equation (\ref{eq:p1vsEnergy}) in Appendix.\ref{appendix:RayTracingStretcher}.
\begin{equation}
    \textbf{p}_1 = \textbf{k}_1+\frac{\delta E}{E_0}\left(\textbf{k}_0-\textbf{k}_1+\frac{\left(\textbf{k}_0-\textbf{k}_1\right)\cdot \textbf{k}_0}{\textbf{k}_1\cdot \textbf{n}_0}\textbf{n}_0\right) .
\end{equation}

To derive the expression for $\textbf{p}_2$, define $\textbf{K}_1$ and $\textbf{K}_2$ to be the wave-vector corresponding to the propagation direction $\textbf{k}_1$, $\textbf{k}_2$, and $\textbf{P}_1$ and $\textbf{P}_2$ for $\textbf{p}_1$ and $\textbf{p}_2$. 
Then 
\begin{gather}
    \textbf{P}_1=\textbf{K}_1+\delta\left(\textbf{K}_1\right) , \\
    \textbf{P}_2=\textbf{K}_2+\delta\left(\textbf{K}_2\right) .
\end{gather}
According to equation (\ref{eq:deltaK1}), 
\begin{gather}
    \delta\left(\textbf{K}_1\right)=\delta\left(\textbf{K}_0\right)+\frac{\left(\textbf{K}_0-\textbf{K}_1\right)\cdot\delta\left(\textbf{K}_0\right)}{\textbf{K}_1\cdot \textbf{n}_0}\textbf{n}_0 , \\
    \delta\left(\textbf{K}_2\right)=\delta\left(\textbf{K}_1\right)+\frac{\left(\textbf{K}_1-\textbf{K}_2\right)\cdot\delta\left(\textbf{K}_1\right)}{\textbf{K}_2\cdot \textbf{n}_1}\textbf{n}_1  .
\end{gather}
Besides, the following definitions hold:
\begin{gather}
    \textbf{k}_i = \textbf{K}_i |\textbf{K}_i|^{-1},~~~~i=0,1,2,\\
    \delta\left(\textbf{K}_0\right) = \textbf{k}_0 \hbar^{-1}\delta E.
\end{gather}
Together, one can show that
\begin{multline}
    \delta \left( \textbf{K}_2 \right) = \frac{ \delta E}{\hbar} \bigg( \textbf{k}_0 - \frac{\left( \textbf{k}_1 + \textbf{k}_2 \right) \cdot \textbf{k}_0}{\textbf{k}_2 \cdot \textbf{n}_1} \textbf{n}_1 \\
    +\frac{\left(\textbf{k}_0-\textbf{k}_1\right)\cdot \textbf{k}_0}{\textbf{k}_1\cdot \textbf{n}_0}\left(\textbf{n}_0-\frac{\left(\textbf{k}_1+\textbf{k}_2\right)\cdot \textbf{n}_0}{\textbf{k}_2\cdot \textbf{n}_1}\textbf{n}_1\right)\bigg) .
\end{multline}
Because 
\begin{equation}
    \textbf{p}_2 = -\frac{\textbf{K}_2+\delta\left(\textbf{K}_2\right)}{\left|\textbf{K}_0\right|+\left|\delta\left(\textbf{K}_0\right)\right|} .
\end{equation}
The linear approximation with respect to $\delta E$ is:
\begin{align}
    \textbf{p}_2 = & ~ \textbf{k}_2-\frac{\delta E}{E_0}\Bigg(\textbf{k}_2+\textbf{k}_0-\frac{\left(\textbf{k}_1+\textbf{k}_2\right)\cdot \textbf{k}_0}{\textbf{k}_2\cdot \textbf{n}_1}\textbf{n}_1  \nonumber \\
     & - \frac{\left(\textbf{k}_0-\textbf{k}_1\right)\cdot \textbf{k}_0}{\textbf{k}_1\cdot \textbf{n}_0} \left(\textbf{n}_0 - \frac{\left(\textbf{k}_1+\textbf{k}_2\right)\cdot \textbf{n}_0}{\textbf{k}_2\cdot \textbf{n}_1}\textbf{n}_1\right) \Bigg) .
\end{align}

Now, with the help of \emph{Mathematica}, one can combine the concrete expressions of $\textbf{n}_0$, $\textbf{n}_1$, $\textbf{k}_0$, $\textbf{k}_1$, $\textbf{k}_2$ and equation (\ref{eq:RayTracingCompressor1}) to get the explicit dependence of the path length on the photon energy
\begin{align}
    \delta(L) &=-\frac{l_1\times8c_1^2}{\sin^2{\left(\alpha_1+\theta_2\right)}}\times\frac{\delta E}{E_0} \nonumber \\
    & \ \ \ \ \ \ \ +\frac{\left(2f-z_1-z_2\right)\times c_2^2}{4\sin^2{\left(\alpha_2-\theta_2\right)}\sin^2{\left(\alpha_1+\theta_2\right)}}\times\frac{\delta E}{E_0}   \label{eq:compressorPathLengthGeneral}
\end{align}
where
\begin{align}
    c_1&=\sin{\left(\theta_1\right)}\sin{\left(\alpha_1-\theta_1+\theta_2\right)} ,\\
    c_2 &= 2\cos{\left(\alpha_1-\theta_1+2\theta_2\right)}\sin(\theta_1-\alpha_2) \nonumber \\
    & \ \ \ \ \ \ \ + \sin{(\alpha_1+\alpha_2)} 
    +\sin{\left(\alpha_1-\alpha_2-2\theta_1\right)} \nonumber \\
    &\ \ \ \ \ \ \ \ \ \ - 2\sin{\left(\alpha_1-\alpha_2\right)} .
\end{align}
In the numerical simulation, $z_1 = z_2 = 10~\text{cm} \ll f=31.5~\text{m}$.
Therefore, we assume that $2f - z_1 - z_2 \approx 2f$.
In the numerical simulation, $\theta = 12.0167\degree$, $\theta_1 = 12.0185\degree$ and $\theta_2 = 12.0182\degree$.
Numerical calculation confirms that these tiny differences do not have significant confluence on the coefficient of $\delta E / E_0$ in equation (\ref{eq:compressorPathLengthGeneral}).
Therefore, in the body text, we assume that $\theta \approx \theta_1 \approx \theta_2$.
In this way, the equation (\ref{eq:compressorPathLengthGeneral}) reduces to equation (\ref{eq:compressorPL}).

\end{document}